\documentclass[
        a4paper,
        authorcolumns,
        cleveref,
        USenglish,
    ]{lipics-v2021}
\usepackage[utf8]{inputenc}
\nolinenumbers

\usepackage[]{auxlib}
\usepackage{bbold}

\usepackage[subtle]{savetrees}

\usepackage{todonotes}

\usepackage{algorithm}
\usepackage[noend]{algorithmic}

\usepackage{multirow}

\newcommand*{\ph}{p^\wedge}
\newcommand*{\pc}{p^\vee}
\newcommand*{\vph}{{\mathbf{p}}^\wedge}
\newcommand*{\vpc}{{\mathbf{p}}^\vee}
\newcommand*{\vc}{{\mathbf{c}}}

\newcommand*{\vv}{{\mathbf{v}}}
\newcommand*{\vw}{{\mathbf{w}}}

\newcommand*{\wtI}{\widetilde{\mathcal{I}}}

\newcommand*{\alg}{\mathrm{ALG}}
\newcommand*{\opt}{\mathrm{OPT}}

\newcommand*{\com}{\mathsf{C}}

\newcommand*{\0}{\mathbf{0}}
\newcommand*{\1}{\mathbf{1}}

\newcommand*{\NP}{\mathsf{NP}}
\newcommand*{\PP}{\mathsf{P}}

\newcommand*{\leqp}{\leq_{\mathsf{p}}}

\def\SLPB/{\textsc{SLTB}}
\def\SLPBtc/{\textsc{SLTB}$_{\textrm{TC}}$}
\def\SLPBm/{\textsc{SLTB}$_{\textrm{M}}$}

\usepackage{thmtools}
\usepackage{thm-restate}

\newenvironment{proofof}[1]{{\vspace*{5pt} \noindent\bf Proof of #1:  }}{\hfill\rule{2mm}{2mm}\vspace*{5pt}}

\makeatletter
\newcommand{\Rom}[1]{\uppercase\expandafter{\romannumeral #1\relax}}
\newcommand{\rom}[1]{\lowercase\expandafter{\romannumeral #1\relax}}
\makeatother

\graphicspath{{Figures/}}

\usepackage[
        numbers,
        sort&compress,
    ]{natbib}
\title{Scheduling with a Limited Testing Budget}
\subtitle{Tight Results for the Offline and Oblivious Settings}

\titlerunning{Scheduling with a Limited Testing Budget}


\author{Christoph Damerius}{Universität Hamburg, Germany}{christoph.damerius@uni-hamburg.de}{}{}
\author{Peter Kling}{Universität Hamburg, Germany}{peter.kling@uni-hamburg.de}{https://orcid.org/0000-0003-0000-8689}{}
\author{Minming Li}{City University of Hong Kong, Hong Kong}{minming.li@cityu.edu.hk}{}{}
\author{Chenyang Xu}{East China Normal University, China}{cyxu@sei.ecnu.edu.cn}{}{}
\author{Ruilong Zhang}{University at Buffalo, USA}{ruilongz@buffalo.edu}{}{}

\authorrunning{C.\ Damerius, P.\ Kling, M.\ Li, C.\ Xu, R.\ Zhang}

\Copyright{Christoph Damerius,  Peter Kling,  Minming Li,  Chenyang Xu,  and Ruilong Zhang}


\EventEditors{John Q. Open and Joan R. Access}
\EventNoEds{2}
\EventLongTitle{42nd Conference on Very Important Topics (CVIT 2016)}
\EventShortTitle{CVIT 2016}
\EventAcronym{CVIT}
\EventYear{2016}
\EventDate{December 24--27, 2016}
\EventLocation{Little Whinging, United Kingdom}
\EventLogo{}
\SeriesVolume{42}
\ArticleNo{23}

\begin{document}
\maketitle

\begin{abstract}
Scheduling with testing falls under the umbrella of the research on optimization with explorable uncertainty.
In this model, each job has an upper limit on its processing time that can be decreased to a lower limit (possibly unknown)
by some preliminary action (testing).
Recently, D{\"{u}}rr et al. \cite{DBLP:journals/algorithmica/DurrEMM20} has studied a setting where testing a job takes a unit time, and the goal is to minimize total completion time or makespan on a single machine.
In this paper, we extend their problem to the budget setting in which each test consumes a job-specific cost, and we require that the total testing cost cannot exceed a given budget.
We consider the offline variant (the lower processing time is known) and the oblivious variant (the lower processing time is unknown) and aim to minimize the total completion time or makespan on a single machine.

For the total completion time objective, we show NP-hardness and derive a PTAS for the offline variant based on a novel LP rounding
scheme.
We give a $(4+\epsilon)$-competitive algorithm for the oblivious variant based on a framework inspired by the worst-case lower-bound instance.
For the makespan objective, we give an FPTAS for the offline variant
and
a $(2+\epsilon)$-competitive algorithm for the oblivious variant.
Our algorithms for the oblivious variants under both objectives run in time $\mathcal{O}(poly(n/\epsilon))$.
Lastly, we show that our results are essentially optimal by providing matching lower bounds.

\end{abstract}

\clearpage
\setcounter{page}{1}
\resetlinenumber[1]
\section{Introduction}

With increased interest in applying scheduling algorithms to solve real-life problems, many models and methods have been addressing the uncertainty in the scheduling community.
Several elegant models that capture uncertainty have been studied in the past two decades, most of which fall under the umbrella of the research on robust optimization~\cite{kasperski2016robust,DBLP:journals/corr/abs-2212-07682,DBLP:journals/tcs/KasperskiZ11,yu1993complexity} or stochastic optimization~\cite{DBLP:journals/corr/abs-2208-13696,DBLP:journals/mor/GuptaMUX20,DBLP:journals/mp/GuptaKNS22,DBLP:journals/mor/Gupta0NS21}.
In those settings, the uncertainty is usually described by the input.
In robust optimization, the input consists of several scenarios, while the input is sampled from a known distribution in stochastic optimization.
In some practical cases, we can gain additional information about the input
by paying extra costs, e.g., money, time, energy, memory, etc.
This model is also known as
{\em explorable uncertainty}, which aims to study the trade-offs between the exploration cost and the quality of a solution.

An intriguing scheduling model for explorable uncertainty was proposed by D{\"{u}}rr et al. \cite{DBLP:journals/algorithmica/DurrEMM20} under the name of \emph{scheduling with testing}.
In their model, before executing a job, one can invest some time to \emph{test} that job, potentially reducing its processing time. 
A practical use case is code optimization, where we could either simply run programs/codes (jobs) as they are or preprocess them through a code optimizer to hopefully improve their execution times.

Their model considers the test cost as the time spent by the machine, which is certainly important and captures many applications stated in~\cite{DBLP:journals/algorithmica/DurrEMM20}.
However, it may fail to describe some scenarios.
For example, in the code optimization problem, the code optimizer may be an expert who might need to be employed by other companies.
This situation is usually faced by cloud computing companies~\cite{kudtarkar2010cost,cardoso2017embedded}, which accept some tasks and want to assign them to servers.
They can employ experts to optimize some time-intensive tasks to speed up the execution.
In this way, the server can finish more tasks, thus creating more profit for the company.
After optimizing, the experts return the optimized tasks to the cloud computing company, and the company can start to assign tasks to servers.
Thus, optimizing does not use servers' time.
Different tasks may require a different amount of effort from the expert to optimize and therefore needs different cost.
The company has a fixed budget and aims to select some tasks to optimize (test) such that the total processing time of tasks is minimized.

Informally, we consider a natural variant of the model proposed in~\cite{DBLP:journals/algorithmica/DurrEMM20}, in which we are given a set of $n$ jobs and a total budget $B$ for testing.
Each job $j$ has an upper limit on the processing time $\ph_j$ and testing cost $c_j$.
After testing, the processing time of job $j$ decreases to a lower limit $\pc_j$, which is possibly hidden for the algorithms.
We refer to the model as the {\em offline} version
if $\pc_j$ is known by the algorithm for all jobs $j$; otherwise, it is called {\em oblivious} version.
The paper considers two objectives: the total completion time objective and the
makespan
objective, which are two well-studied objectives for scheduling problems in the literature~\cite{HallSchulz:97:Scheduling-to-minimize,BrunoCoffman:74:Scheduling-independent,Karp:72:Reducibility-among,DuLeung:89:Complexity-of-scheduling}. 
The formal definition of our problem is stated in \cref{sec:model}. 

Note that the offline version of the model in
D{\"{u}}rr et al. \cite{DBLP:journals/algorithmica/DurrEMM20} is easy, even if testing a job $j$ requires a job-specific amount of time $t_j$.
Testing a job is then beneficial if $p_j^\wedge-p_j^\vee>t_j$.
In contrast, we show that the offline version of our budgeted variant of the problem is NP-hard, assuming each job takes a job-specific amount of budget to be tested.
We study both the offline and the oblivious settings.
Further, we differentiate between the uniform cost variant, where each job takes one unit
of budget to be tested, and a non-uniform variant, where the testing cost is job-specific.

\subsection{Our Contributions}

The paper studies the problem of Scheduling with a Limited Testing Budget (SLTB) under both the total completion time minimization objective (\SLPBtc/) and the makespan minimization objective (\SLPBm/). For both objectives, we further distinguish the offline and oblivious settings. 



Our main results are summarized in \cref{table:results}. 
For the objective of total completion time minimization, in the offline setting, we show that the problem is NP-hard even when all the lower processing times are $0$ by a reduction from the \textsc{Partition} problem, and then give a PTAS.
The PTAS is derived based on a novel LP rounding scheme.
Further, we find that there exists an FPTAS if all the jobs share the same lower processing time. For the oblivious setting, we give a $(4+\epsilon)$-competitive deterministic algorithm for any $\epsilon$ (we use the concept of the competitive ratio following the previous work~\cite{DBLP:journals/algorithmica/DurrEMM20}). 
The ratio is almost tight since we prove that no deterministic algorithm has a competitive ratio strictly better than $4$.
For the objective of makespan minimization, the main results are derived based on a connection between our problem and the classical 0-1 knapsack problem. We prove that the offline setting is NP-hard and admits an FPTAS, while for the oblivious setting, an almost tight competitive ratio of $2+\epsilon$ can be obtained.




\begin{table}[htp]
\begin{tabular}{cccccc}
\hline
\multicolumn{1}{l}{}       & \multicolumn{1}{l}{} & UB (\SLPBtc/) & LB (\SLPBtc/) & UB (\SLPBm/) & LB (\SLPBm/)             \\ \hline
\multirow{2}{*}{Offline}   & $\vpc\in\R_{\geq 0}^{n}$                   & PTAS {\scriptsize(Thm. \ref{thm:offline:ptas})}      & \multirow{2}{*}{NP-C {\scriptsize(Thm. \ref{thm:offline:l1norm})}} & \multirow{2}{*}{FPTAS {\scriptsize(Cor. \ref{cor:makespan:offline:FPTAS})}} & \multirow{2}{*}{NP-C {\scriptsize(Cor. \ref{cor:makespan:offline:hardness})}} \\ 
                           & $\vpc \in \R_{\geq 0} \cdot \1$            & FPTAS {\scriptsize(Cor. \ref{cor:online:l1norm:pc})} &                          \\ \hline
{Oblivious} & ---          & $4+\epsilon$ {\scriptsize(Thm. \ref{thm:online:l1norm:ratio})} & {4 {\scriptsize(Thm. \ref{thm:online:l1norm:hardness})}} & $2+\epsilon$ {\scriptsize(Thm. \ref{thm:online:makespan:non-uniform:ratio})}           & {2 {\scriptsize(Thm. \ref{thm:online:makespan:hardness})}}       \\ \hline 
\end{tabular}
\vspace{1em}
\caption{The summary of our results. The vector $\vpc:=(\pc_1,\ldots,\pc_n)$ is the lower processing time vector, and $\vpc \in \R_{\geq 0} \cdot \1$ means that all the entries of the vector share the same value. $\epsilon$ is an arbitrary positive parameter.}
\label{table:results}
\end{table}
\vspace{-2.7em}
%
%
%
%
\paragraph{Paper Organization} We first state some useful notation in~\cref{sec:model}, and then give an overview of our techniques in~\cref{sec:techn}.
In the remaining part of the main body (\cref{sec:offline_slpbtc}),
we describe a PTAS for the offline SLTB with the total completion time objective, the most interesting and technical part of our work. Due to space limitations, the proofs are deferred to \Cref{sec:omitted_details_slpbtc}. We leave the details for our other results for \cref{sec:hardness,sec:makespan,sec:oblivious}.

\subsection{Related Work}

\paragraph{Explorable Uncertainty} Scheduling with testing falls under the umbrella of the research on optimization with explorable uncertainty, where some additional information can be obtained through queries. 
The model under the stochastic setting can be traced back to Weitzman’s Pandora’s Box problem~\cite{weitzman1978optimal} and it remains an active research area up to the present~\cite{DBLP:conf/ipco/GuptaN13,DBLP:conf/icml/GergatsouliT22}.
The model under the adversarial setting was first coined by Kahan~\cite{DBLP:conf/stoc/Kahan91} to study the number of queries necessary to obtain an element set's median value. 
So far, many optimization problems have been considered in this setting, e.g. caching~\cite{DBLP:conf/vldb/OlstonW00}, geometric tasks~\cite{DBLP:journals/mst/BruceHKR05}, minimum spanning tree~\cite{DBLP:conf/stacs/HoffmannEKMR08,DBLP:journals/siamcomp/MegowMS17}, knapsack~\cite{DBLP:journals/cor/GoerigkGISS15} and so on. 

\paragraph{Scheduling with Testing} 
The problem of scheduling with testing was first coined by D{\"{u}}rr et al. \cite{DBLP:journals/algorithmica/DurrEMM20}.
They consider a model where each testing operation requires one unit of time and mainly investigate non-preemptive schedules on a single machine to minimize the total completion time or makespan.
Since the offline version of the problem (algorithms know the lower processing time of each job) is trivial, they mainly consider the online version.
They present a $2$-competitive deterministic algorithm for total completion time minimization while the deterministic lower bound is $1.8546$.
They also gave a $1.7453$-competitive randomized algorithm while the randomized lower bound is $1.6257$.
For makespan minimization, they give a $1.618$-competitive deterministic algorithm and show that it is optimal for the deterministic setting.
They also present a $4/3$-competitive randomized algorithm and show that it is optimal.

Later, Albers and Eckl \cite{DBLP:conf/waoa/AlbersE20} consider the non-uniform testing case where the testing time depends on the job.
They investigate the single-machine preemptive and non-preemptive scheduling to minimize the total completion time or makespan.
The offline version of this problem is still trivial, so they mainly consider the oblivious version.
They present a $4$-competitive deterministic algorithm for total completion time minimization and a $3.3794$-competitive randomized algorithm.
If preemption is allowed, the deterministic ratio can be further improved to $3.2361$.
All lower bounds are the same as in the uniform testing case.
For makespan minimization, they extend the algorithm proposed in \cite{DBLP:journals/algorithmica/DurrEMM20} and show that the approximation can be preserved in the non-uniform testing case.

Scheduling with testing on identical machines is also considered in the literature~\cite{DBLP:conf/wads/AlbersE21}.
The authors mainly consider the makespan minimization in both non-preemptive and preemptive settings.
They look into the non-uniform testing case.
For the preemptive setting, they present a $2$ competitive algorithm which is essentially optimal.
For the non-preemptive setting, they give a $3.1016$-competitive algorithm for the general testing case, and the ratio can be improved to $3$ when each test requires one unit of time.
Later, Gong et al. \cite{gong2022improved} improved the non-preemptive ratios to $2.9513$ and $2.8081$ for non-uniform and uniform testing cases, respectively.




\section{Preliminaries}
\label{sec:model}

 An \emph{instance} to Scheduling with a Limited Testing Budget (\SLPB/) is a 5-tuple $\mathcal{I}=(J,\vph,\vpc,\mathbf{c},B)$.
$J=[n]$ denotes a \emph{set of $n$ jobs}.
Each job $j$ has an \emph{upper limit on the processing time} $\ph_j\in\mathbb{R}_{\ge 0}$, a \emph{lower processing time} $\pc_j \in\intcc{0,\ph_j}$ and a \emph{testing cost} $c_j\in\mathbb{R}_{\ge 0}$.
These parameters are collected in the \emph{lower and upper limit processing time vectors} $\vpc$ and $\vph$, respectively, and
a \emph{vector of testing costs} $\mathbf{c}$.
Additionally, a total amount of \emph{budget} $B\in\mathbb{R}_{\ge 0}$ is given.


Each job $j$ can be executed either in a tested or untested state.
When job $j$ is tested, $j$ will take $\pc_j$ time to process; otherwise, it requires $\ph_j$ time.
If a job is tested, it consumes $c_j$ budget; otherwise, no budget is consumed.

We consider offline and oblivious versions.
For the offline version, the algorithm knows the complete instance $\mathcal{I}$.
For the oblivious version, the lower processing time vector $\vpc$ is hidden from the algorithm, and the remaining information of the instance is known a priori.

In this work, we only consider non-preemptive and, w.l.o.g., gapless schedules on a single machine.
Once a job starts executing, other jobs cannot be processed until the current job is finished.
Thus, a schedule corresponds to a specific ordering of jobs.
We define $I\coloneqq [n]$ to be the \emph{set of positions}.
The job in position $i\in I$ will be the $i^{th}$ job executed in the schedule.

A \emph{schedule} $S=(\sigma,J_\vee)$ for an instance $\mathcal{I}=(J,\vph,\vpc,\mathbf{c},B)$ is defined by a \emph{job order} $\sigma$ and a \emph{testing job set} $J_{\vee}\subseteq J$.
The job order $\sigma: J\rightarrow I$ is a bijective function that describes the order in which the jobs are processed
(i.e., job $j$ is the $\sigma(j)$-th processed job in the non-preemptive schedule).
The testing job set $J_\vee\subseteq J$ represents a set of jobs to test with $\sum_{j\in J_\vee}c_j\le B$.

Given a schedule $S$, we can indicate whether a job is tested using a \emph{set of types} $T\coloneqq \set{\vee,\wedge}$.
We say that $j$ is \emph{of type $\vee$,$\wedge$} if it is \emph{tested}, \emph{untested}, respectively.
If $S$ schedules a job $j$ of type $t$ into position $i$, we also say that \emph{position $i$ is of type $t$}.
For a schedule $S=(\sigma,J_\vee)$ and a job $j$, let the \emph{type $t_S(j)$ of $j$ in $S$} be $\vee$ if $j\in J_\vee$ and $\wedge$ otherwise.
Denote by $C_j\coloneqq \sum_{j'\in J, \sigma(j')\le \sigma(j)} p_{j'}^{t_S(j')}$ the \emph{completion time} of job $j$ in schedule $S$.
The total completion time is the sum of all completion times, i.e., $\sum_{j\in J}C_j$, and the \emph{makespan} is the maximum completion time among all jobs, i.e., $\max_{j\in J}\{C_j\}$.


Given a testing job set $J_\vee$, the optimal ordering of the jobs is easy to determine.
The ordering is relevant for the total completion time minimization but not for the makespan minimization.
It is a well-known fact that the SPT rule (shortest processing time first) orders
the jobs optimally for total completion time minimization.
The processing times are in our case $p_j^\vee$ if job $j$ is tested and $p_j^\wedge$ otherwise.
Thus, an optimal schedule can be easily constructed from an optimal testing job set $J_\vee$.

\section{Overview of Techniques}
\label{sec:techn}

In this section, we focus on the total completion time minimization and give technical overviews for the offline model and the oblivious model.


\subsection{Offline \SLPB/ under Total Completion Time Minimization}
\label{sec:techn:tc:offline}

For offline \SLPBtc/, we mainly show the following theorem.
The NP-hardness is proved via a reduction from the \textsc{Partition} problem. Due to space limitations, we defer the proof to \cref{sec:hardness} and focus on introducing the high-level ideas of our PTAS, the most interesting and technical part of this paper. 

\begin{restatable}{theorem}{offlinecomplete}
\label{thm:offline:l1norm}
The offline \SLPBtc/ problem is NP-hard even when the lower processing time of each job is $0$, and admits a PTAS.
\end{restatable}
Our algorithm is based on an integer linear programming (ILP) formulation for offline \SLPBtc/.
The ILP contains variables $x_{j,i,t}$ that dictate whether job $j\in J$ should
be scheduled in position $i\in I$ of type $t\in T$.
(See \cref{subsec:ILP_formulation}
for the exact definition of this ILP.)
The ILP is conceptually similar to the classical matching ILP on bipartite graphs~\cite{DBLP:books/daglib/0069809}, with jobs and positions
representing the two disjoint independent sets of the bipartition.
A matching would then describe an assignment of jobs to positions.
However, there are two main differences.
First, we have two variables per pair of job and position (distinguished by the type $t\in T$).
This translates to each job-position pair having two edges that connect them in the (multi-)graph.
Second, the total cost of jobs tested is restricted by some budget $B$.
This causes a dependency when selecting edges in the graph.

Our approach combines a rounding scheme of the ILP with an exploitation of the cost structure of the problem.
We relax the ILP to an LP by allowing the variables $x_{j,i,t}$ to take on fractional values between $0$ and $1$.
We start with an optimal LP solution and then continue with our rounding scheme, which consists of two phases.
In the first phase, we round the solution such that all fractional variables correspond to the edges
of a single cycle in the graph mentioned above.
These variables are hard to round directly without overusing the budget.
Here we start the second rounding phase.
We relax some of the constraints in the LP to be able to continue the rounding process.
Specifically, we allow certain positions to schedule two jobs (we call these positions \emph{crowded}).
We end up with an integral (but invalid) solution that has some crowded positions.
Then, we "decrowd" these positions by moving their jobs to nearby positions (shifting the position of some other jobs one up), and show that we can bound the cost of moving a job this way in terms of its current contribution to the overall cost. Observing that moving a job from position $i$ to position $i'$ (note that positions are counted from right to left) increases that job's contribution by a factor of $i'/i$, if a crowded position lies far to the right ($i$ is small), we cannot afford to move one of its jobs too far away. For example, in the extreme case that the rightmost position is crowded (i.e., $i = 1$), even the smallest possible move of one of its jobs to the second-rightmost position (i.e., $i' = 2$) already doubles that job's contribution. Thus, our algorithm tries to avoid producing crowded positions that lie too far to the right (at small positions).

To this end, the rounding process in this phase is specifically tailored to control where crowded positions can appear in the integral solution. We look at the $f(\epsilon) = 2/\epsilon + 1$ smallest (rightmost) positions that appear on the current path (representing fractional variables), and select one of them (let's call it the cut-position) to cut the path into two halves. This is done such that each half contains $1/\epsilon$ many of the smallest positions on the current path. By shifting workload along each of these two halves, we can make one of them integral. This integral half gives us $1/\epsilon$ positions that are not crowded, while the cut-position might have become crowded (as might any future cut-position in the remaining fractional path). Because we cut somewhere in the $f(\epsilon)$ rightmost positions of the path, we can show that for each crowded position, there are many positions further to the right of the schedule that are not crowded (this is basically what our charging argument formalizes). In the end, this allows us to prove that no job is moved too far from its original position (relative to its original position), keeping the cost increase due to such moves small.

\subsection{Oblivious \SLPB/ under Total Completion Time Minimization}
\label{sec:techn:tc:online}

For the oblivious model where the lower processing time vector $\vpc$ is hidden, we show that $(4+\epsilon)$ approximation can be obtained, and further, prove that the ratio is the best possible.

\begin{restatable}{theorem}{onlinecomplete}
\label{thm:online:l1norm}
For oblivious \SLPBtc/ and any $\epsilon>0$, there exists a deterministic algorithm with a competitive ratio of $(4+\epsilon)$, while no deterministic algorithm can obtain a competitive ratio strictly smaller than $4$.  
\end{restatable}

We start by considering the oblivious uniform \SLPBtc/ problem to build some intuition. 
The uniform case limits the number of tested jobs, i.e., we can test at most $k$ jobs. 
Clearly, for the worst-case analysis, we can assume that each job $j$ tested by our algorithm has $ \pc_j=\ph_j $; that is, we exhaust the budget, but no job's processing time gets reduced. 
In contrast, for all the jobs tested by an optimal solution, their processing times can be reduced to $0$. 
Thus, from this perspective, regardless of which jobs we test, our total completion time remains unchanged, but the optimum depends on our tested jobs because the adversary can only let the job $j$ that is not tested by our algorithm have $\pc_j=0$.

Then we find that the oblivious uniform \SLPBtc/ problem is essentially equivalent to the following optimization problem: given a set of jobs $J$ with $\vph$ and $\vpc = \0$, the goal is to select $k$ jobs such that the minimum total completion time obtained by testing at most $k$ unselected jobs is maximized. 
The selected jobs can be viewed as the jobs tested by our algorithm, while the minimum total completion time obtained by testing  unselected jobs is the optimum of oblivious uniform \SLPBtc/. 
When our objective value is fixed, a larger optimum implies a better competitive ratio. 
For this much easier problem, it is easy to see that the best strategy is selecting the $k$ jobs with the largest upper processing time, which is the set of jobs that would be tested by an optimal solution of \SLPBtc/ instance $\cI=(J,\vph,\vpc=\0, \vc=\1,k)$.

We build on the above argument to give the algorithm for the non-uniform case $\cI=(J, \vph,\vpc, \vc, B)$. 
The basic idea is constructing an auxiliary instance $\wtI:=(J,\vph,\widetilde{\vpc}=\0,\vc, B)$, solving the instance optimally or approximately, and returning the obtained solution. 
Use $\alg(\cdot)$ and $\opt (\cdot)$ to denote the objective values obtained by our algorithm and an optimal solution of an input instance, respectively. 
By the theorem proved in the offline model, we have $\alg(\wtI) \leq (1+\epsilon) \opt (\wtI)$ for any $\epsilon>0$. 
In the analysis, we show that our objective value can be split into two parts: $\alg(\cI) \leq 2\alg(\wtI) + 2\opt (\cI)$, and therefore, due to $\opt(\wtI)\leq \opt(\cI)$, a competitive ratio of $(4+2\epsilon)$ can be proved.

The lower bound is shown by a hard instance $\cI=(J,\vph=\1,\vpc,\vc=\1, B=\frac{n}{2})$, where the adversary always lets our tested jobs have lower processing time $1$ and the processing time of any other job be $0$. 
Apparently, any deterministic algorithm's objective value is $n(n+1)/2$, while an optimal solution can achieve a total completion time of $n(n+2)/8$, which implies a lower bound of $4$.
\footnote{
Since in the worst-case, the upper and lower processing times of jobs tested by the algorithm are equal,
it does not help if the algorithm can be adaptive, i.e., change its testing strategy based on such an information.
}

\subsection{\SLPBm/ under Makespan Minimization}
\label{sec:techn:m}

\begin{theorem}\label{thm:makespan_main}
The offline \SLPBm/ problem is NP-hard and admits an FPTAS, while for oblivious \SLPBm/, an almost tight competitive ratio of $2+\epsilon$ can be obtained (for any $\epsilon>0$).
\end{theorem}

The offline \SLPB/ problem under makespan minimization is closely related to the classical 0-1 knapsack problem.
The classical 0-1 knapsack problem aims to select a subset of items such that (\rom{1}) the total weight of the selected items does not exceed a given capacity; (\rom{2}) the total value of the selected items is maximized.
To see the connection, consider the testing cost of each job as the weight of each item and the profit of testing a job ($\ph_j-\pc_j$) as the value of an item.
Then we build on the algorithmic idea of the knapsack dynamic programming and design an FPTAS for the offline setting.

We use the same framework as the total completion time minimization model for the oblivious setting and obtain a $(2+\epsilon)$-competitive algorithm. 
The ratio becomes better here since, for the makespan objective, we have $\alg(\cI) \leq \alg(\wtI) + \opt(\cI)$, saving a factor of $2$. 
The lower bound proof is also based on the same hard instance $\cI=(J,\vph=\1,\vpc,\vc=\1, B=n/2)$. Any deterministic algorithm's makespan is $n$ while the optimum is $n/2$, giving a lower bound of $2$. 

\section{Offline Setting for SLTB under Total Completion Time Minimization}
\label{sec:offline_slpbtc}

This section considers the Scheduling with a Limited Testing Budget problem under total completion time minimization (\SLPBtc/) in the offline setting and aims to show the following theorem.

\begin{restatable}{theorem}{offlineptas}
     \label{thm:offline:ptas}
    There exists a PTAS for \SLPBtc/.
\end{restatable}

For convenience, we refer to a problem instance as a pair $\cI = (J, B)$, dropping the processing time and cost vectors $\vpc$, $\vph$, and $\mathbf{c}$ (which we assume to be implicitly given).
Moreover, in this section, we consider the job positions $I = [n]$ \emph{in reverse order} to simplify the calculations.
That is, a job $j$ scheduled in position $i \in I$ is processed as the \emph{$i$-th last} job.


\subsection{ILP Formulation and Fixations}%
\label{subsec:ILP_formulation}

We start by introducing our ILP formulation of the \SLPBtc/ problem and defining the term \emph{fixation} of a (relaxed) instance of our ILP.
Such fixations allow us to formally fix the values of certain variables in the (relaxed) ILP when analyzing our algorithm.

\vspace{0.5em}
\noindent \textbf{ILP Formulation.}
Our ILP has indicator variables $x_{j,i,t}$ that are $1$ if job $j$ is scheduled at position $i$ of type $t$ and $0$ otherwise.
The \emph{contribution} of such a job to the total completion time is\footnote{%
    Remember that we consider the position in \emph{reverse} order.
    Thus, the job at position $i$ is the $i$-th last job.
} $i \cdot p_j^t$.
We have constraints to ensure that each of the $n$ positions schedules one job, that each job is scheduled once, and that the cost of tested jobs do not exceed the budget.
The equivalence between ILP solutions and \SLPBtc/ schedules is formalized in \cref{lem:schedule_ilp_equivalence2}.

	Consider an instance $\cI = (J, B)$ of the \SLPBtc/ problem.
	We define an ILP $ILP_{\mathcal{I}}$, with the variables $x_{j,i,t}$ for each job $j\in J$, position $i\in I$ and type $t\in T$.
	{\small\begin{align*}
		\min &~~~~\sum_{\mathclap{j\in J, i\in I, t\in T}} ~~~~\,i \cdot p_j^t \cdot x_{j,i,t}\\
		\mathrm{s.t. }&\;\;\:~\sum_{j\in J, t\in T} ~~x_{j,i,t}~~\,\:~~\:\;= 1~~~\forall i\in I~~(1)~~~~\,~\,\sum_{i\in I, t\in T} x_{j,i,t} = 1 ~~~~~~~\:\forall j\in J~~~~~~~~~~~~~~~~~~(2)
		\\
		&\sum_{j\in J, i\in I} c_j x_{j,i,\vee}\le B~~~~~~~~~~~~~~~\,~~~~~~(3) ~~~~~~~~~~~~~\,\,~\;x_{j,i,t}\in\set{0,1}~\forall j\in J, i\in I, t\in T~~(4)
	\end{align*}}
	For $ILP_{\mathcal{I}}$ with variable set $X_{\mathcal{I}}\coloneqq \set{x_{j,i,t} | j\in J, i\in I, t\in T}$,
	a \emph{solution} $x:X_{\mathcal{I}}\rightarrow\mathbb{R}$ assigns each variable in $X_{\mathcal{I}}$ a value.
	Solution $x$ is called \emph{valid} if it satisfies the four constraints and \emph{invalid} otherwise.
	For a (possibly invalid) solution $x$ for $\mathcal{I}$ we define its \emph{cost} as $C_{\mathcal{I}}(x)\coloneqq \sum_{j\in J, i\in I, t\in T} i\cdot p_j^t \cdot x_{j,i,t}$ and its \emph{budget use} $B_{\mathcal{I}}(x)\coloneqq \sum_{j\in J,i\in I} c_j x_{j,i,\vee}$ (we omit $\mathcal{I}$ from $C_{\mathcal{I}}$ and $B$ if it is clear from the context).
	We refer to the different constraints as (1) \emph{position constraints}, (2) \emph{job constraints}, (3) \emph{budget constraint}, and (4) \emph{integrality constraints}.


\begin{restatable}{lemma}{scheduleilpequivalenceto}
    \label{lem:schedule_ilp_equivalence2}
	Let $\mathcal{I}=(J,B)$ be an instance for \SLPBtc/.
	For each valid solution $x$ to $ILP_{\mathcal{I}}$ there exists a schedule $S$ for $\mathcal{I}$ with $C(S)=C(x)$ and vice versa.
	Each can be computed from the other in polynomial time.
\end{restatable}

\noindent {\bf Relaxation and Fixations.}
Our algorithm and analysis use relaxed variants of $ILP_{\cI}$ that \emph{fix} certain ILP variables (indicating that, e.g., certain jobs must be tested).
It also keeps track of \emph{crowded} positions, in which our algorithm may (temporarily) schedule two jobs (violating the position constraints).
We introduce the notion of a \emph{fixation} $\cF$ to formally define these relaxed variants $LP_{\cI,\cF}$ of $ILP_{\cI}$.
\footnote{
    Our PTAS will enumerate through a polynomial number of fixations,
    and solve the problem for each one of them.
    The approximation guarantee is then derived for the fixation that
    is consistent with the optimal solution.
}

\begin{definition}%
\label{def:fixation2}
A \emph{fixation} $\mathcal{F}=(J(\mathcal{F}),X(\mathcal{F}),I^C(\mathcal{F}))$ of an \SLPBtc/ instance $\cI$ consists of:
\begin{enumerate}[nosep]
\item a set of \emph{tested} jobs $J(\mathcal{F})\subseteq J$,
\item a set of \emph{fully-fixed} variables $X(\mathcal{F})\subseteq X_{\mathcal{I}}$ where $j\notin J(\mathcal{F})$ for all $x_{j,i,t}\in X(\mathcal{F})$, and
\item a set of \emph{crowded positions} $I^C(\mathcal{F})\subseteq I$.
\end{enumerate}
\end{definition}

For a set operator $\circ\in\set{\cup,\cap,\setminus}$ and a set of positions $\bar{I}\subseteq I$, we use the notation $\mathcal{F} \circ \bar{I}\coloneqq (J(\mathcal{F}),X(\mathcal{F}),I^C(\mathcal{F}) \circ \bar{I})$ to express the change to the crowded positions of $\cF$.

Given a fixation $\cF$, we define the following relaxed variant $LP_{\cI,\cF}$ of $ILP_{\cI}$:
\begin{enumerate}[topsep=0.5\topsep]
\item For each $x_{j,i,t}\in X_{\cI}$ we relax the integrality constraint to $0\le x_{j,i,t}\le 1$ (\emph{unit constraints}).
\item For each $x_{j,i,t}\in X(\mathcal{F})$, we add the constraint $x_{j,i,t}=1$ (\emph{fully-fixed constraints}).
\item For each $j\in J(\mathcal{F})$, we add the constraint $\sum_{i\in I}{x_{j,i,\vee}}=1$ (\emph{tested job constraints}).
\item For each $i \in I^C(\cF)$, we relax the position constraint to $\sum_{j\in J,t\in T} x_{j,i,t}\in\set{0,1,2}$.
\end{enumerate}
The resulting LP is given in \ref{LP_fix} in the appendix.

\subsection{Graph-theoretic Perspective \& Paths}%
\label{subsec:graphtheoretic_perspective}

Consider an \SLPBtc/ instance $\cI = (J, B)$ with fixation $\cF$ and a (fractional) solution $x$ to $LP_{\cI,\cF}$.
The main building block of our algorithm is a rounding scheme based on the following graph-interpretation of $\cI$ and corresponding paths based on the current solution $x$:
\begin{definition}
The \emph{instance graph} $G_{\cI} \coloneqq (J \cupdot I, E)$ is a bipartite multi-graph between the jobs $J$ and positions $I$ with exactly two edges between any pair $j \in J$ and $i \in I$.
We identify the edge set $E$ with the variable set $X_{\cI}$ and refer to a variable $x_{j,i,t} \in E = X_{\cI}$ also as an \emph{edge of type $t \in \set{\vee,\wedge}$ between $j$ and $i$}.
\end{definition}
\begin{definition}
A \emph{path $P$ in solution $x$} is a weighted path from $i_s \in I$ (\emph{start position}) to $i_e \in I$ (\emph{end position}) in $G_{\cI}$, where the weight of an edge $x_{j,i,t} \in P$ is its value in $x$.
$P$ is called \emph{integral} if all its weights are integral and \emph{fractional} if they are all (strictly) fractional.

Nodes and edges in $P$ must be pairwise distinct, except for possibly equal start and end positions $i_s = i_e$, in which case we refer to $P$ also as a \emph{cycle}.
We define $J(P)$ as the path's set of jobs, $I(P)$ as its set of positions, and $K(P) \coloneqq I(P) \setminus \set{i_s, i_e}$.
Moreover, $X(P)$ is the sequence of edges/variables from start to end position in $P$.
We say the $i$-th edge in $X(P)$ is even/odd if $i$ is even/odd, such that $P$ reaches $j \in J(P)$ via an odd edge $x_j^O$ and leaves $j$ via an even edge $x_j^E$.
We similarly use $t_j^O$ and $t_j^E$ to denote the type of $x_j^O$ and $x_j^E$, respectively.
\end{definition}

Next, we define \emph{shift operations}, which move workload along paths by increasing the volume of one job at any position $i \in I(P)$ while decreasing the volume of another job at $i$.
\begin{definition}%
\label{def:shiftoperation}
A \emph{$\delta$-shift} of a path $P$ in $x$ decreases the value of all odd edges (variables) of $P$ by $\delta$ and increases the value of all even edges (variables) of $P$ by $\delta$.
\end{definition}

\begin{figure}
\begin{subfigure}{0.5\linewidth}
\centering\includegraphics[width=0.7\linewidth]{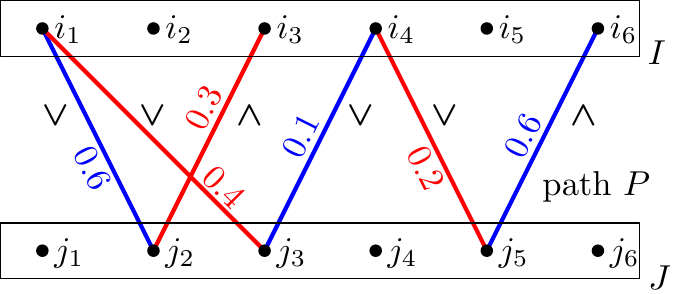}
\caption{Path $P$ with start and end positions $i_3$ and $i_6$.}
\end{subfigure}
\begin{subfigure}{0.5\linewidth}
\centering\includegraphics[width=0.7\linewidth]{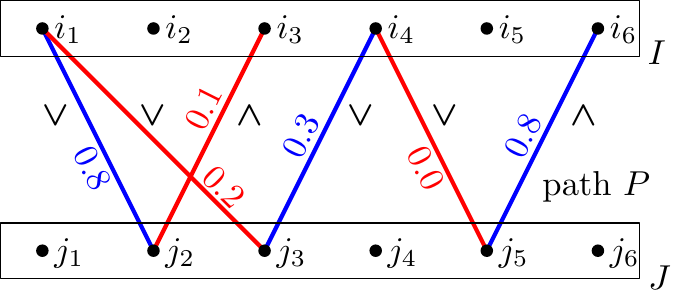}
\caption{Same path $P$ after the shift.}
\end{subfigure}
\caption{%
    A path $P$ in a solution $x$ before and after a shift by $\delta = 0.2$.
    Edges are labeled with their type ($\vee$ or $\wedge$) and weight from a given solution $x$.
    Odd edges are red, and even edges are blue.
    The budget rate computes as $\Delta(P)=c_{j_2} (1-0)+c_{j_3} (1-1)+c_{j_5}(0-1)=c_{j_2}-c_{j_5}$.
}%
\label{fig:path_shift}
\end{figure}

Shift operations (see \cref{fig:path_shift}) change the budget use $B(x)$ at a path-dependent (positive or negative) \emph{budget rate} (defined below) and might create crowded positions.
Our algorithm's first two phases (\cref{subsec:first_phase:elibutone, subsec:second_phase:roundblocking}) carefully pair shift operations such that performing paired shifts does not increase the budget and does not create too many crowded positions.


Define the \emph{budget rate} of a path $P$ to be $\Delta(P) \coloneqq \sum_{j\in J(P)} c_j \cdot(\mathbb{1}_{|t_j^E=\vee}-\mathbb{1}_{|t_j^O=\vee})$. Let $P$ be a path in a solution $x$ for $LP_{\cI,\cF}$ without crowded positions (i.e., $I^C(\cF) = \varnothing$).
$P$ is called \emph{$y$-alternating} (or simply \emph{alternating}) if all odd edges have weight $y$ and all even edges have weight $1-y$. \Cref{lem:alternation_property2} below formalizes the effect of a $\delta$-shift in terms of the path's budget rate.
Since our analysis can be restricted to paths with very specific, alternating edge values, we also formalize such \emph{alternating paths} and show how they are affected by $\delta$-shifts (see also \cref{fig:multiple_path_shift}).


\begin{restatable}{lemma}{alternationpropertyto}
    \label{lem:alternation_property2}
Let $P$ be a path in a solution $x$ for $LP_{\cI,\cF}$ without crowded positions (i.e., $I^C(\cF) = \varnothing$).
Shifting $P$ in $x$ by $\delta$ yields a (possibly invalid) solution $\tilde{x}$ with $B(\tilde{x}) = B(x) - \delta \cdot \Delta(P)$.
If $P$ is $y$-alternating in $x$, then it is $(y-\delta)$-alternating in $\tilde{x}$.
\end{restatable}


\subsection{First Phase: Eliminating all but one cycle}%
\label{subsec:first_phase:elibutone}

Consider an optimal valid solution $x$ to $LP_{\mathcal{I},\mathcal{F}}$ without crowded positions (i.e., $I^C(\cF) = \varnothing$).
\Cref{lem:merging_paths2} below is our main tool for rounding fractional variables in $x$.
Consider a set of variables that form a fractional path in $x$.
Essentially, we want to use a shift operation from \cref{def:shiftoperation} on such a path to make some of its variables integral.
If such a shift increases the budget use $B(x)$ (rendering the solution invalid), we can suitably shift a second path (possibly using a negative $\delta$) in parallel to ensure that the budget use $B(x)$ does not increase.

\begin{figure}
\begin{subfigure}{0.5\linewidth}
\centering\includegraphics[width=0.7\linewidth]{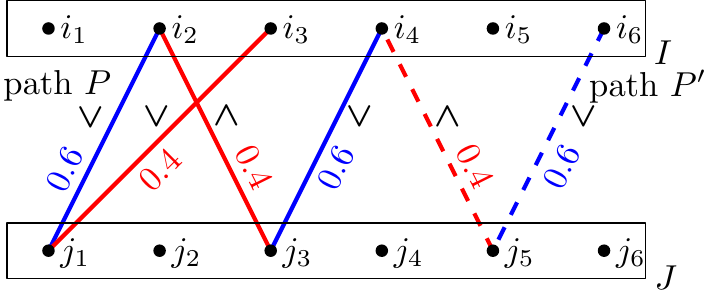}
\caption{Alternating paths $P$ (solid) and $P'$ (dashed).}
\end{subfigure}
\hfill
\begin{subfigure}{0.5\linewidth}
\centering\includegraphics[width=0.7\linewidth]{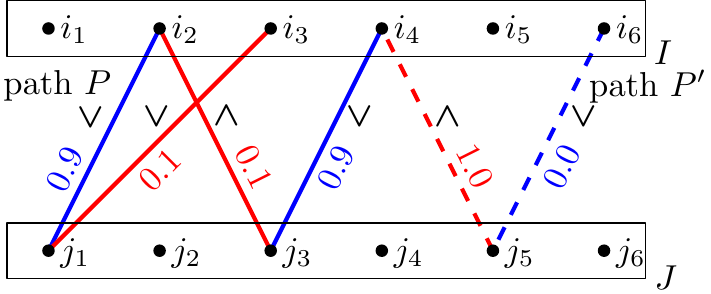}
\caption{Same paths by $\delta = 0.3$ and $\delta' = 0.6$, respectively.}
\end{subfigure}
\caption{%
    Illustration of \Cref{lem:merging_paths2} for alternating paths $P = (i_3,\dots,i_4)$ (solid) and $P' = (i_4,\dots,i_6)$ (dashed) with budget rates  $\Delta(P) = c_{j_3} \coloneqq 2$ and $\Delta(P') = c_{j_5} \coloneqq 1$.
    Shifting $P$ by $\delta = 0.3$ and $P'$ by $\delta' = 0.6$ keeps the budget use constant.
    $P$ and $P'$ stay alternating, $P'$ becomes integral, and $i_3,i_4,i_6$ (the start/end positions) violate the position constraints after these shifts.
}%
\label{fig:multiple_path_shift}
\end{figure}

Such shifts might also cause the violation of the position constraints at the path's start and end positions.
We keep track of such violations by adding those positions to the crowded position set $I^C(\cF)$ of the fixation $\cF$.
\Cref{lem:merging_paths2} formalizes this approach (see also \cref{fig:multiple_path_shift}).

\begin{restatable}{lemma}{mergingpathsto}
    \label{lem:merging_paths2}
Consider $x$ a valid solution for $LP_{\cI,\cF}$.
Let $P$ be a fractional path in $x$ with $\Delta(P)=0$ or
$P,P'$ be two fractional paths in $x$ with $X(P) \ne X(P')$ and $\Delta(P), \Delta(P')\ne 0$.
We can efficiently shift $P$ (and $P'$, if existing) in $x$ to yield a valid solution $\tilde{x}$ for $LP_{\cI,\tilde{\cF}}$ with:
\begin{enumerate}[nosep]
\item $C(\tilde{x})\le C(x)$ and $B(\tilde{x})=B(x)$
\item $\tilde{\mathcal{F}}=\mathcal{F}\cup I'$, where $I'$ is the set of all start and end positions
    of non-cyclic paths involved.
\item $\tilde{x}$ contains more integral variables than $x$.
\end{enumerate}
\end{restatable}

\Cref{lem:merging_paths2} allows us to shift along general paths (instead of cycles) at the cost of creating crowded positions.
We rely on this in \cref{subsec:second_phase:roundblocking} and deal with the crowded positions in \cref{subsec:third_phase_shifting}.
However, to keep the number of crowded positions small and reduce their impact on the final solution, we apply \cref{lem:merging_paths2} on cycles for as long as possible.
This avoids the creation of crowded positions since shifts along cycles cannot change the net workload at any position.
Indeed, note that given any node in a path $P$ that is incident to a fractional edge must have a second fractional edge, or it would violate its job/position constraint.
This allows us to complete any fractional path to a cycle.
Thus, we can keep applying \cref{lem:merging_paths2} to \emph{cycles} (not creating crowded positions) until there is at most one cycle with a non-zero budget rate left (a \emph{blocking cycle}).
Let $x$ be a solution to $LP_{\mathcal{I},\mathcal{F}}$.
A path $P$ of $x$ is called \emph{critical} if $X(P)=\set{x_{j,i,t}\in X_{\mathcal{I}} | x_{j,i,t}\in\intoo{0,1}}$.
A \emph{blocking cycle} of $x$ is a critical cycle $P$ with $\Delta(P)\ne 0$. \Cref{lem:find_decomposition2} formalizes the idea above.



\begin{restatable}{lemma}{finddecompositionto}
    \label{lem:find_decomposition2}
Let $\mathcal{I}$ be an instance, $\mathcal{F}$ be a fixation with $I^C(\mathcal{F})=\varnothing$,
and $x$ be a valid optimal solution to $LP_{\mathcal{I},\mathcal{F}}$.
Then we can compute in polynomial time a valid optimal solution $\tilde{x}$ to
$LP_{\mathcal{I},\mathcal{F}}$
such that
all variables in $\tilde{x}$ are integral, or
we find a blocking cycle of $\tilde{x}$. Further, if $P$ is a blocking cycle of $x$, then $P$ is alternating. 
\end{restatable}




\subsection{Second Phase: Rounding the blocking cycle}%
\label{subsec:second_phase:roundblocking}

Assume that we used \Cref{lem:find_decomposition2} to compute a blocking cycle $P$ for a solution $x$ to $LP_{\mathcal{I},\mathcal{F}}$ with $I^C(\mathcal{F})=\varnothing$.
Because $P$ is critical, all fractional variables are in $X(P)$.
Also, since $\Delta(P)\ne 0$, applying
\Cref{lem:merging_paths2} directly is impossible.
Instead, we cut up $P$ repeatedly into two paths $P_1,P_2$, and then use \Cref{lem:merging_paths2} on these paths (see \Cref{alg:offline:round_paths}).
Cutting is done by selecting any position $i\in K(P)$, and separating the path at $i$: $P_1$ will be the
path starting at the start position of $P$ and end at $i$. $P_2$ will be the path starting at $i$ and ending at the end position of $P$. We abbreviate this operation by $P_1,P_2\gets \textsc{Cut}(P,i)$.
The drawback of this approach is that \Cref{lem:merging_paths2} does not guarantee that the resulting solutions still fulfill the position constraints of the start/end positions of $P_1,P_2$, respectively.
That is why we add them to $I^C(\mathcal{F})$ in the process.

Algorithm $\textsc{RepeatedCut}$ starts with a solution $\tilde{x}$ and a critical path $\tilde{P}$. It cuts $\tilde{P}$ at some position $i\in K(\tilde{P})$ that is selected by a procedure $\textsc{SelectCutPosition}$ (which is described later in \cref{alg:offline:select_cut_position} in the next subsection).
The algorithm then applies \Cref{lem:merging_paths2} to the two resulting paths, making at least one of them integral (as guaranteed by \Cref{lem:alternation_property2}).
After that, $\tilde{x}$ and $\tilde{P}$ are updated accordingly.
$\textsc{RepeatedCut}$ finishes when $\abs{K(\tilde{P})}=0$
(and therefore $\tilde{P}$ cannot be cut into two paths anymore).
In such a case, $\textsc{RepeatedCut}$ will reschedule that job to obtain an integral solution.
It is also possible that no path remains after the application of \Cref{lem:merging_paths2}.
For such a case, $\tilde{x}$ is already integral.
Thus, in both cases, the resulting integral solution $\tilde{x}$ is returned.

\begin{algorithm}[htb]\small
\caption{$\textsc{RepeatedCut}$}
\label{alg:offline:round_paths}
\begin{algorithmic}[1]
\REQUIRE A valid solution $x$ for $LP_{\mathcal{I},\mathcal{F}}$ with $I^C(\mathcal{F})=\varnothing$, a blocking cycle $P$ of $x$.
\STATE {$\tilde{P},\tilde{x}\gets P,x$}
\WHILE{$\Delta(\tilde{P})\ne 0$}
    \IF {$K(\tilde{P})=\varnothing$}
        \STATE {$j\gets $ unique job in $J(\tilde{P})$; $i_1,i_2\gets $ remaining two positions in $I(\tilde{P})$}
        \STATE {In $\tilde{x}$, reschedule $j$ into position $\min(i_1,i_2)$ of type $\vee$ if $t_j^O=t_j^E=\vee$ and $\wedge$ else}
        \RETURN {$\tilde{x}$}
    \ENDIF
	\STATE{$i\gets \textsc{SelectCutPosition}(\tilde{P})$}
    \STATE{$P_1,P_2\gets \textsc{Cut}(P,i)$}
	\STATE{Apply \Cref{lem:merging_paths2} to $P_1,P_2$ in $\tilde{x}$, changing $\tilde{x}$ accordingly}
	\STATE {\textbf{if} both paths became integral \textbf{then} \textbf{return} $\tilde{x}$}
	\STATE{$\tilde{P}\gets $ the remaining fractional path}
\ENDWHILE
\STATE{Apply \Cref{lem:merging_paths2} to $\tilde{P}$ in $\tilde{x}$, changing $\tilde{x}$ accordingly}
\RETURN $\tilde{x}$
\end{algorithmic}
\end{algorithm}

In the following, we make statements about the state of the variables involved in the execution of $\textsc{RepeatedCut}$ at the beginning of an iteration of its \textbf{while}-loop.
Consider the state of $\textsc{RepeatedCut}$ (called on path $P$) at the beginning of the $l$'th iteration of
    the \textbf{while}-loop ($l\ge 1$).
    We denote by $I^C_l$ the start/end position of $P$ together with all positions selected by \textsc{SelectCutPosition} so far, and $I^C_*$ the start/end position of $P$ together with all positions selected by \textsc{SelectCutPosition} throughout the algorithm.
    Similarly, denote by $\tilde{x}_l$,$\tilde{P}_l$ the values
    of $\tilde{x}$,$\tilde{P}$ at that point, respectively, and $\tilde{x}_{*}$ for the returned solution by $\textsc{RepeatedCut}$.
    Denote $\tilde{\mathcal{F}}_l\coloneqq (J(\mathcal{F}),X(\mathcal{F}),I^C_{l})$ and
    $\tilde{\mathcal{F}}_*\coloneqq (J(\mathcal{F}),X(\mathcal{F}),I^C_*)$.


\begin{restatable}{lemma}{algofflineloopinvariant}
     \label{lem:alg_offline:loop_invariant}
    The following is a loop invariant of $\textsc{RepeatedCut}$ for iteration $l\ge 1$:
    $\tilde{x}_l$ is a valid solution for $LP_{\mathcal{I},\tilde{\mathcal{F}}_l}$ and $\tilde{P}_l$ is a critical fractional alternating path in $\tilde{x}_l$, of which the start and end positions are in $I^C_l$.
    Also, $\tilde{x}_*$ is an integral valid solution for $LP_{\mathcal{I},\tilde{\mathcal{F}}_*}$.
\end{restatable}

Based on \Cref{lem:alg_offline:loop_invariant}, we can analyze the objective obtained by \textsc{RepeatedCut}:


\begin{restatable}{lemma}{algrepeatedcutproperties}
    \label{lem:alg_repeated_cut_properties}
	Consider an application of $\textsc{RepeatedCut}$ on solution $x$ for $LP_{\mathcal{I},\mathcal{F}}$ and a blocking cycle $P$.
	It returns in polynomial time a solution $\tilde{x}$ with
    $C(\tilde{x})\le C(x)+Z$, where $Z$ is either $0$ or the contribution of job $j$ rescheduled by \textsc{RepeatedCut} in line 5 and $j\notin J(\tilde{\mathcal{F}}_*)$.
\end{restatable}


\subsection{Third Phase: Dealing with crowded positions}
\label{subsec:third_phase_shifting}

\Cref{lem:alg_repeated_cut_properties} guarantees that applying the algorithm $\textsc{RepeatedCut}$ will
return us an integral solution. However, that solution is valid for $LP_{\mathcal{I},\mathcal{F}}$ where $I^C(\mathcal{F})$ still contains some positions.
Some of these positions may schedule two jobs, which makes this schedule not valid for $ILP_{\mathcal{I}}$.
Our general strategy in this subsection is to move the jobs such that the cost of the solution does not increase
too much. In \Cref{obs:alg_offline_moving_jobs}, we move each job to a new position and bound the
cost created by that operation.


\begin{restatable}{observation}{algofflinemovingjobs}
    \label{obs:alg_offline_moving_jobs}
    Let $x$ be an integral solution to $LP_{\mathcal{I},\mathcal{F}}$ for some fixation $\mathcal{F}$.
    Consider a job $j\in J$ that is scheduled in position $i\in I$ of type $t\in T$.
    Then rescheduling $j$ into position $i'$, i.e., setting $x_{j,i,t}\gets 0$ and $x_{j,i',t}\gets 1$ produces a (possibly invalid) solution $\tilde{x}$, in which the contribution of $j$ increases by a factor of $i'/i$ compared to $x$.
\end{restatable}


As mentioned above, there are still some positions that schedule two jobs.
To obtain an integral valid solution for $ILP_{\mathcal{I}}$, we have to move the jobs in the schedule
to new positions, such that there is exactly one job per position scheduled.
We want to use \Cref{obs:alg_offline_moving_jobs} to bound the increase in contribution for each job moved this way.
Generally, we move the jobs as follows:
For a position $i$ that schedules two jobs $j,j'$, we (arbitrarily) distribute $j,j'$ among positions
$i,i+1$, thereby moving all jobs from positions $i+1,\dots,n$ to one higher position.
Following this strategy, jobs in higher positions may get moved multiple times.

To bound the contribution in terms of \Cref{obs:alg_offline_moving_jobs}, we set up a charging scheme:
Each position with two jobs scheduled
should be charged to a distinct set of $1/\epsilon$ smaller positions that schedule one job, where $\epsilon$
is the accuracy parameter of our algorithm ($1/\epsilon\in\mathbb{N}$).
In the following, we will always use the following function $\textsc{SelectCutPosition}$ for $\textsc{RepeatedCut}$:

\begin{algorithm}[htb]\small
\caption{$\textsc{SelectCutPosition}$}
\label{alg:offline:select_cut_position}
\begin{algorithmic}[1]
\REQUIRE A path $\tilde{P}$ with $\abs{K(\tilde{P})}\ge 1$
\ENSURE A position $i\in K(\tilde{P})$
\IF{$\abs{K(\tilde{P})}\ge 2/\epsilon+1$}
	\STATE{$I'\gets$ the smallest $2/\epsilon+1$ positions in $K(\tilde{P})$}
	\RETURN {the position that appears as $(1/\epsilon+1)$-st position in $\tilde{P}$ of the positions in $I'$}
\ELSE
	\RETURN{any position in $K(\tilde{P})$}
\ENDIF
\end{algorithmic}
\end{algorithm}

We care about two properties of the position selected by $\textsc{SelectCutPosition}$.
First, when we cut $\tilde{P}$ into $P_1,P_2$ in line 8 of $\textsc{RepeatedCut}$,
$K(P_1)$,$K(P_2)$ should each contain at least $1/\epsilon$ positions.
This way, whichever of these paths becomes integral, there will be $1/\epsilon$ positions
that will never be selected by $\textsc{SelectCutPosition}$ in the future.
This is important for our charging scheme to have enough positions to charge to.
Second, we specifically care about the selected positions being the smallest positions that appear in
$K(\tilde{P})$.
This essentially allows us to charge each position with two jobs scheduled exclusively to smaller positions, independent of which of the two paths becomes integral.

We represent the charging scheme using a \emph{charging function}
(formally defined in \Cref{def:support_function}).
Essentially, for a set of positions $\bar{I}\subseteq I$, it charges each position
in $\bar{I}$ to its distinct $1/\epsilon$ many smaller positions.

\begin{definition}
    \label{def:support_function}
    Let $x$ be a solution to $LP_{\mathcal{I},\mathcal{F}}$ for an instance $\mathcal{I}$ and a fixation $\mathcal{F}$.
    Let $P$ be a critical path in $x$.
    For a set $\bar{I}\subseteq I$, a \emph{charging function} for $\bar{I}$ is a function
    $\mu: \bar{I}\rightarrow \cP(I\setminus \bar{I})$ such that for all $i\in \bar{I}$:
    (1) $\abs{\mu(i)}=1/\epsilon$, (2) $\forall i'\in \mu(i): i'<i$ and (3) $\forall i'\in \bar{I}: \mu(i)\cap \mu(i')=\varnothing$.
    
    Consider the $l$'th iteration of the \textbf{while}-loop in $\textsc{RepeatedCut}$.
    We define the \emph{charging set}
    $I^+_l\coloneqq \bar{I}\cup I^C_l$, where
    $\bar{I}$ contains the smallest $2/\epsilon+1$ positions in $K(\tilde{P}_l)$
    (or all of them, if $\abs{K(\tilde{P}_l)}<2/\epsilon+1$).
    Similarly, define $I^+_*\coloneqq I^C_*$.
\end{definition}

\begin{figure}
\mbox{}\hfill
\begin{subfigure}{0.45\linewidth}
\centering\includegraphics[width=0.7\linewidth]{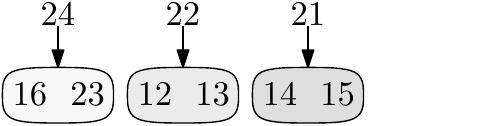}
\caption{Illustration of $\mu_l$ (e.g., $\mu_l(24) = \set{16, 23}$).}
\end{subfigure}
\hfill
\begin{subfigure}{0.5\linewidth}
\centering\includegraphics[width=0.62\linewidth]{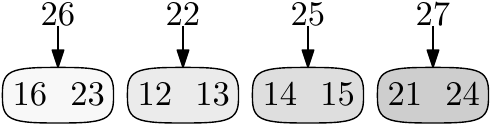}
\caption{Illustration of $\mu_{l+1}$ (e.g., $\mu_{l+1}(26) = \set{16, 23}$).}
\end{subfigure}
\hfill\mbox{}

\medskip

\begin{subfigure}{\linewidth}
\centering\includegraphics[width=0.65\linewidth]{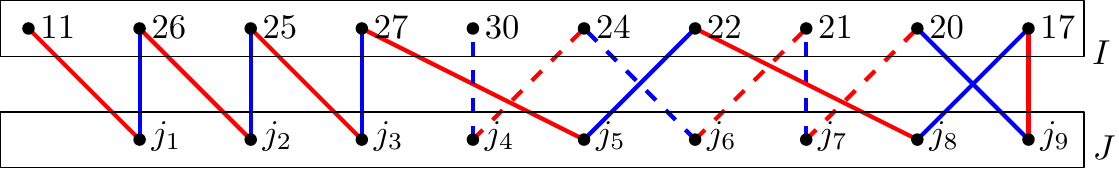}
\caption{%
    Path $\tilde{P}_l$ (solid \emph{and} dashed), and path $\tilde{P}_{l+1}$ (solid) after one iteration of the \textbf{while}-loop in \textsc{RepeatedCut}.
}
\end{subfigure}
\caption{%
    An update step of \Cref{lem:mu_function_update} for $\epsilon=1/2$.
    The inner positions of $\tilde{P}_l$ are $K(\tilde{P}_l)=\set{17,20,21,22,24,25,26,27}$.
    Its $2/\epsilon+1 = 5$ smallest positions appear in order $22,17,20,21,24$.
    $\tilde{P}_l$ is cut at the $(1/\epsilon+1) = 3$-rd of these positions ($20$) into two paths, which are then shifted such that the dashed one becomes integral and the solid one becomes $\tilde{P}_{l+1}$.
    $I^{+}_{l+1}$ loses positions $21$ and $24$ compared to $I^+_l$, as these positions belonged to the dashed path, which became integral. $I^{+}_{l+1}$ now consists of all remaining $2/\epsilon+1 = 5$ inner positions $K(\tilde{P}_{l+1})=\set{17,22,25,26,27}$.
    We set $\mu_{l+1}(25)=\mu_{l}(21)$, $\mu_{l+1}(26)=\mu_{l}(24)$, and $\mu_{l+1}(27)=\set{21,24}$ (the lost positions from $I^{+}_l$).
}
\label{fig:mu_function_update}
\end{figure}
\Cref{lem:mu_function_update} shows
how to obtain a charging function
$\mu_{*}: I^+_*\rightarrow \cP(I\setminus I^+_*)$
from a charging function
$\mu_1: I^+_1\rightarrow \cP(I\setminus I^+_1)$.
We do this by updating the charging function with
every iteration of $\textsc{RepeatedCut}$'s loop.
\Cref{fig:mu_function_update} exemplifies the update of the charging function.

\begin{restatable}{lemma}{mufunctionupdate}
\label{lem:mu_function_update}
    If there exists a charging function $\mu_1: I^+_1\rightarrow \cP(I\setminus I^+_1)$, then there
    also exists a charging function $\mu_{*}: I^+_*\rightarrow \cP(I\setminus I^+_*)$.
\end{restatable}

We now use \Cref{obs:alg_offline_moving_jobs} together with a charging function (of which we assume the existence for now)
on a solution $x$ for $LP_{\mathcal{I},\mathcal{F}}$ produced by $\textsc{RepeatedCut}$ to find
a solution for $ILP_{\mathcal{I}}$ with not too much more cost.
\Cref{lem:find_ilp_solution} will allow us to produce such a solution.


\begin{restatable}{lemma}{findilpsolution}
    \label{lem:find_ilp_solution}
    Let $x$ be a solution returned by $\textsc{RepeatedCut}$ for $LP_{\mathcal{I},\mathcal{F}}$,
    and let $\mu_*$ be a charging function for $I^+_*$. Then we can find a valid solution $\tilde{x}$ in polynomial time for $ILP_{\mathcal{I}}$ such that the contribution of each job increases by a factor of at most $(1+\epsilon)$ compared to $x$.
\end{restatable}


Consider a solution $x$ returned by $\textsc{RepeatedCut}$ for $LP_{\mathcal{I},\mathcal{F}}$.
To be able to apply \Cref{lem:find_ilp_solution} and find a solution for $ILP_{\mathcal{I}}$,
we need to make sure that we can find a charging function $\mu_1$ for $I^+_1$.
Furthermore, we still need to bound the contribution created by the job $j$ in line 5 of \textsc{RepeatedCut} as of \Cref{lem:alg_repeated_cut_properties}.
To do this, we choose a proper fixation $\mathcal{F}^*$ 
and show that a charging function can then be derived.

\begin{definition}
    \label{def:fixation_brute_force}
    Let $x^*$ be an optimal solution to
    $ILP_{\mathcal{I}}$ for an instance $\mathcal{I}$.
    For $i\in I$, let further $j_i\in J$ and $t_i\in T$ such that $x^*_{j_i,i,t_i}=1$.
    Using $M\coloneqq (2/\epsilon+1)/\epsilon$,
    we define the fixation $\mathcal{F}^*$ by
    \begin{align*}
    	X(\mathcal{F}^*)=\set{x_{j_i,i,t_i}|i\in [M]}
    	&~&
    	J(\mathcal{F}^*)=\set{j\in J|p_j^\wedge>\min_{i\in[M]}{p_{j_i}^{t_i}}}
    	&~&
    	I^C(\mathcal{F}^*)=\varnothing
    \end{align*}
\end{definition}


\begin{restatable}{lemma}{initialsupportfunction}
     \label{obs:initial_support_function}
    Let $x$ be a solution returned by $\textsc{RepeatedCut}$ for $LP_{\mathcal{I},\mathcal{F}^*}$.
    Then there exists a charging function $\mu_1$ for $I^+_1$.
\end{restatable}

 Essentially, we brute-force which jobs
will be scheduled in the last few positions.
This will make sure that these positions are not in $I^+_1$, and as such can be used
for the charging function $\mu_1$.
Assuming that we brute-forced correctly, an optimal solution will also test all jobs with a larger upper processing time than any of the brute-forced jobs.
This is because an optimal solution will always schedule
in order of increasing processing times.
Finally, we can piece together all of the above lemmas and prove the main theorem (\cref{thm:offline:ptas}).
The detailed proof can be found in~\cref{sec:omitted_details_slpbtc}.

\section{Conclusion}%
\label{sec:conclusion}

We initiated the study of Scheduling with a Limited Testing Budget, where we have a limited budget for testing jobs to potentially decrease their processing time.
We provided NP-hardness results, a PTAS, as well as tight bounds for a semi-online (oblivious) setting.

Our results open promising avenues for future research.
For the setting where we minimize the total completion time, it remains open whether NP-hardness holds for uniform testing cost.
Also, while our LP-rounding-based PTAS achieves the best possible approximation, it remains open whether there is a faster, combinatorial algorithm.
Another natural direction would be to consider the case of multiple machines.

Another exciting direction is the following \emph{bipartite matching with testing} problem that generalizes our problem, arising from the graph-theoretic perspective in \cref{subsec:graphtheoretic_perspective}:
Consider a bipartite graph $G \coloneqq (L \cup R, E)$ in which each edge $e \in E$ has a cost $c_e$ that can be reduced to $\check{c}_e$ via a testing operation.
Given the possibility to test edges before adding them to the matching, we seek a min-cost perfect matching that respects a given testing budget.

\section*{Acknowledgment}
We thank the anonymous reviewers for their many insightful comments and suggestions.
Chenyang Xu was supported in part by Science and Technology Innovation 2030 –``The Next Generation of Artificial Intelligence" Major Project No.2018AAA0100900.
Ruilong Zhang was supported by NSF grant CCF-1844890.

\clearpage

\printbibliography
\clearpage
\appendix
\crefalias{section}{appendix}

\makeatletter
\@mkboth{APPENDIX}{APPENDIX}
\makeatother

\section{Omitted Details from \cref{sec:offline_slpbtc} (PTAS for \SLPBtc/)}%
\label{sec:omitted_details_slpbtc}

\scheduleilpequivalenceto*
\begin{proof}
From a solution $x$ to $ILP_{\mathcal{I}}$, we construct a schedule $S=(\sigma,J_{\vee})$ by setting
$\sigma(j)\coloneqq \sum_{j\in J,i\in I, t\in T} (n-i+1) \cdot x_{j,i,t}$ and $J_{\vee}=\set{j\in J | \sum_{i\in I}{x_{j,i,\vee}}=1}$.

Consider any job $j\in J$.
Because of $j$'s job constraint and the integrality constraints in $ILP_{\mathcal{I}}$, there exists exactly one variable
$x_{j,i_j,t_j}=1$, and all other variables for job $j$ are zero.
Therefore $\sigma(j)=n-i_j+1$.
Furthermore, because of $i_j$'s position constraint, there cannot be another job $j'$ with $\sigma(j')=\sigma(j)$.
We conclude that $\sigma$ is a bijection.
$J_{\vee}$ is a valid set of tested jobs ($S$ is a schedule) because
\begin{align*}
    \sum_{j\in J_\vee} c_j=\sum_{j\in J, \sum_{i\in I}{x_{j,i,\vee}}=1} c_j=\sum_{j\in J, i\in I} x_{j,i,\vee}=B_{\mathcal{I}}(x)\le B.
\end{align*}

Abbreviating $p_j\coloneqq p_j^{t_S(j)}$, we get
\begin{align*}
    C(S)
    =&\sum_{\bar{j}\in J}{C_{\bar{j}}}
    =\sum_{\bar{j}\in J} \sum_{j\in J, \sigma(j)\le \sigma(\bar{j})} p_j
    =\sum_{j\in J} \sum_{\bar{j}\in J, \sigma(j)\le \sigma(\bar{j})} p_j
    =\sum_{j\in J} p_j \sum_{\bar{j}\in J, \sigma(j)\le \sigma(\bar{j})} 1\\
    =&\sum_{j\in J} p_{j} (n-\sigma(j)+1)
    =\sum_{j\in J} i_j\cdot p_j^{t_j} \cdot x_{j,i_j,t_j}
    =\sum_{j\in J, i\in I, t\in T} i\cdot p_j^t \cdot x_{j,i,t}=C_{\mathcal{I}}(x)
\end{align*}
The converse statement can be analogously obtained.
Obviously, schedules and solutions $x$ can be converted into each other in polynomial time.
\end{proof}

\paragraph{LP Relaxations via Fixations}
In the following, we state the linear program resulting from a fixation $\mathcal{F}$ defined in \Cref{def:fixation2}.
	
\begin{align}
    \mathrm{min. }&\sum_{j\in J, i\in I, t\in T} i\cdot p_j^t \cdot x_{j,i,t}& \tag{\text{LP$_{\cI,\cF}$}} \label{LP_fix} \\
	\mathrm{s.t. }&~~\sum_{j\in J, t\in T} ~~\:x_{j,i,t}\:= 1~~~\forall i\in I\setminus I^C(\mathcal{F})& \sum_{i\in I, t\in T} &x_{j,i,t} \:= 1 ~~~~~\:\forall j\in J
	\notag \\
	&\sum_{j\in J, i\in I, t\in T} x_{j,i,\vee}\le B&0\le \;&x_{j,i,t}\:\le 1~~~~~~\forall j\in J, i\in I, t\in T
	\notag \\
	&~~~~\:\sum_{i\in I}~~~\:\,x_{j,i,\vee}=1~~~\forall j\in J(\mathcal{F})& \sum_{j\in J,t\in T} &x_{j,i,t}\;\in\intcc{0,2}~~\forall i\in I^C(\mathcal{F})\notag \\
	&~~~~~~~~~~~~~~x_{j,i,t}~=1~~\;\forall x_{j,i,t}\in X(\mathcal{F})& \notag
\end{align}

\alternationpropertyto*
\begin{proof}
    We determine how the budget used by $\tilde{x}$ compares to the budget used by $x$.
	\begin{align*}
		B(\tilde{x})
        =&\sum_{\tilde{x}_{j,i,\vee}\in X_{\mathcal{I}}} c_j\cdot \tilde{x}_{j,i,\vee}
		=\sum_{\tilde{x}_{j,i,\vee}\in X_{\mathcal{I}}\setminus X(P)}{c_j\cdot  \tilde{x}_{j,i,\vee}}
        +\sum_{\tilde{x}_{j,i,\vee}\in X(P)}{c_j\cdot  \tilde{x}_{j,i,\vee}}\\
        =&\sum_{\tilde{x}_{j,i,\vee}\in X_{\mathcal{I}}\setminus X(P)}{c_j\cdot  \tilde{x}_{j,i,\vee}}
        +\sum_{j\in J(P)}{c_j\cdot  \left({
        {}
        \tilde{x}_j^O}\cdot \mathbb{1}_{| t_j^O=\vee}+{{}\tilde{x}_j^E}\cdot \mathbb{1}_{| t_j^E=\vee}\right)}\\
        =&\sum_{x_{j,i,\vee}\in X_{\mathcal{I}}\setminus X(P)}{c_j\cdot  x_{j,i,\vee}}
        +\sum_{j\in J(P)}{c_j\cdot  \left(({{}x_{j}^O}+\delta)\cdot \mathbb{1}_{| t_j^O=\vee}+({{}x_j^E}-\delta)\cdot \mathbb{1}_{| t_j^E=\vee}\right)}\\
        =&\sum_{x_{j,i,\vee}\in X_{\mathcal{I}}}{c_j\cdot  x_{j,i,\vee}}
        +\sum_{j\in J(P)}{c_j\cdot  \left(\delta\cdot \mathbb{1}_{| t_j^O=\vee}-\delta\cdot \mathbb{1}_{| t_j^E=\vee}\right)}
        = B(x)-\delta \cdot \Delta(P)
	\end{align*}
    If $P$ is $y$-alternating, $x_j^O=y$ and $x_j^E=1-y$ for any $j\in J(P)$.
    Because of this, $x_j^O$ and $x_j^E$ are the only variables that influence $j$'s job constraint.
    When we shift $P$ by $\delta$, we get that $\tilde{x}_j^O=x_j^O-\delta$ and $\tilde{x}_j^E=x_j^E+\delta$.
    Therefore $P$ will be $(y-\delta)$-alternating in $\tilde{x}$.
\end{proof}

\mergingpathsto*
\begin{proof}
	Let $\delta\in\mathbb{R}$ (to be fixed later).
	Consider first the case that we have only one path $P$ with $\Delta(P)=0$.
	We shift $P$ in $x$ by $r\coloneqq \delta\cdot \Delta(P)$ and show that the resulting solution
	$\tilde{x}$ is a valid solution for $\tilde{L}\coloneqq LP_{\mathcal{I},\tilde{\mathcal{F}}}$ (for proper choice of $\delta$).
	We only modify the variables in $X(P)$.
	Therefore we only have to consider constraints that contain variables $x_{j,i,t}$ where $j\in J(P)$.
	(This implies that we do not have to consider fully-fixed constraints.)

    Job constraints for jobs $j\in J(P)$ remain satisfied, as
    $j$'s odd edge is decreased by $r$ and $j$'s even edge is increased by $r$.
    Assume that a tested job constraint for $j$ exists.
    If $t_j^O=t_j^E$, then this constraint remains satisfied for the same reason above.
    $t_j^O\ne t_j^E$, on the other hand, is impossible: Because of $j$'s valid job constraint,
    $\sum_{i\in I,t\in T} x_{j,i,t}=1$ holds, and since either $t_j^O=\wedge$ or $t_j^E=\wedge$,
    we must have $\sum_{i\in I} x_{j,i,\vee}<1$, which contradicts the tested job constraint for $j$ in the valid solution $x$.

	For the budget constraint, we get that $B(\tilde{x})=B(x)-r\cdot \Delta(P)$
    by \Cref{lem:alternation_property2}. Since $\Delta(P)=0$, we get that $B(\tilde{x})=B(x)$, so the budget constraint remains satisfied.

    Now consider the unit constraints.
    Since all variables in $X(P)$ have values in $\intoo{0,1}$, choosing $\delta$ with small enough magnitude
    will cause these constraints to remain satisfied.
    Shifting is a linear operation, so
	the sign of $\delta$ can be chosen such that $C(\tilde{x})\le C(x)$ after this operation.
	By choosing the magnitude $\delta$ large enough, we can force at least one of the variables in $X(P)$ to become integral, while
	all other integral variables will remain integral.
	This proves the statement of the lemma for a single involved path $P$.

	Now assume that we have two paths $P$, $P'$, and $\Delta(P),\Delta(P')\ne 0$.
	We construct $\tilde{x}$ by shifting $P$ in $x$ by $r\coloneqq \delta\cdot \Delta(P')$ and $P'$ by $r'\coloneqq -\delta\cdot \Delta(P)$.
	The arguments are the same as above, except for the budget constraint.
	Here we combine the two changes to obtain
	\[
        B(\tilde{x})
        =B(x)-r\cdot \Delta(P)-r' \cdot\Delta(P')
        =B(x)-\delta \cdot \Delta(P') \cdot \Delta(P)+\delta \cdot \Delta(P)\cdot \Delta(P')
        =B(x)
	\]
	so the budget constraint is still satisfied.
	The choice of $\delta$ is analogous to above.
    Notice that we can choose $\delta$ such that one of the changed variables becomes integral since $X(P)\ne X(P')$.
	This shows the lemma's statement when two paths are involved.
	We finish the proof by noting that the above procedure can be carried out in polynomial time.
\end{proof}

\finddecompositionto*
\begin{proof}
	If $x$ is integral, the statement is trivially fulfilled.
    Otherwise, we can construct a cycle as follows.
    First, consider all fractional variables contained in $G_{\mathcal{I}}$.
    For each fractional variable $x_{j,i,t}$, there
    must exist two different fractional variables $x_{j,i',t'}$ and $x_{j',i,t''}$.
    This means that all jobs $j$/positions $i$ are incident to two fractional variables in $G_{\mathcal{I}}$.
    This induces a cycle $P_1$ in $G_{\mathcal{I}}$.
    It can be constructed by starting at a fractional variable $x_{j,i,t}$, and iteratively walking to adjacent variables that share a job/position with the previous variable and are also fractional.
    The cycle is completed when a previously used node is repeated.

	Now consider the case where $\Delta(P_1)=0$.
	By \Cref{lem:merging_paths2}, we can find an optimal valid solution $\tilde{x}$ for $L$ that has less fractional variables than $x$.
	Therefore, after iterating this step polynomially many times, all variables become integral or the cycle found satisfies $\Delta(P_1)\ne 0$.

	If all fractional variables in $x$ are in $X(P_1)$, then $P_1$ is a blocking cycle of $x$.
	Otherwise, we aim to construct another cycle $P_2$ with $X(P_2)\ne X(P_1)$ and $\Delta(P_2)\ne 0$.
	There exists another fractional variable $x_{j,i,t}\notin X(P_1)$.
	If there is no such variable such that $j\in J(P_1)$ or $i\in I(P_1)$, then
	we can find another cycle $P_2$ the same way as above, with
	$X(P_2)\cap X(P_1)=\varnothing$ (that cycle cannot loop back into $P_1$).
	Otherwise, choose such a variable $x_{j,i,t}$.
    Using the same procedure as above, we can now construct a second cycle, starting at $x_{j,i,t}$.
    When we select a variable that is already in $X(P_1)$, we can use variables from $X(P_1)$
    to complete the cycle $P_2$. It contains $x_{j,i,t}\notin X(P_1)$, hence $X(P_1)\ne X(P_2)$.
    Otherwise, when we repeat a variable, we constructed a cycle in the same way as $P_1$, with
    $X(P_1)\cap X(P_2)=\varnothing$, so we also get $X(P_1)\ne X(P_2)$.
	If $\Delta(P_2)=0$, then we do the same as for the case above where $\Delta(P_1)=0$, again decreasing the number of fractional variables each step.

	Otherwise, we found two cycles $P_1,P_2$ with $\Delta(P_1),\Delta(P_2)\ne 0$ and $X(P_1)\ne X(P_2)$.
	Again, we apply \Cref{lem:merging_paths2} to obtain an optimal valid solution $\tilde{x}$ for $LP_{\mathcal{I},\mathcal{F}}$ with less fractional variables than $x$.
	Therefore, after repeating this procedure polynomially many times,
	the resulting solution $\tilde{x}$ must either be integral, or we get a blocking cycle of $\tilde{x}$ as described above.

 Now we prove that a blocking cycle $P$ must be alternating.
	By definition, $X(P)$ consists of exactly the variables $x_{j,i,t}\in\intoo{0,1}$ that appear in the valid solution $x$.
    Therefore, for each job $j\in J(P)$ there exist exactly two variables $x_{j}^O,x_{j}^E\in\intoo{0,1}$.
	Because of the job constraint of $j$, there are no other nonzero variables $x_{j,\cdot, \cdot}$ in $x$.
	It follows that $x_j^O+x_j^E=1$.
    The same argument can be made for positions $i\in I(P)$: They have exactly two variables that influence their position constraint.
    Hence, the value of any odd edge in $P$ (say that variable has value $y$) fixes
    the value of all other variables in $X(P)$ to $y$ for odd edges and $1-y$ for even edges.
	As such $P$ must be alternating.
\end{proof}

\algofflineloopinvariant*
\begin{proof}
    We first show that the loop invariant holds before the first execution of the loop.
	Here, $\tilde{x}_1=x$ holds because of line 1. By assumption, $x$ is valid for $LP_{\mathcal{I},\mathcal{F}}$.
	$\tilde{\mathcal{F}}$ differs from $\mathcal{F}$ only in the sense that $I^C(\tilde{\mathcal{F}})=I^C_{0}=\set{i}\supseteq \varnothing=I^C(\mathcal{F})$ where $i$ is $P$'s start/end position.
	Therefore, $LP_{\mathcal{I},\tilde{\mathcal{F}}}$ only relaxes the position constraint at $i$ compared to $LP_{\mathcal{I},\mathcal{F}}$.
	Hence $\tilde{x}_1$ is valid for $LP_{\mathcal{I},\tilde{\mathcal{F}}}$.
	By \Cref{lem:alternation_property2} and \Cref{lem:find_decomposition2}, $P$ is critical and alternating since $P$ is a blocking cycle, so the same holds for $\tilde{P}$ after the assignment in line 1.
	Lastly, the start and end position of $\tilde{P}$ are identical to $i$, which is exactly $I^C_1$. ($\textsc{SelectCutPosition}$ did not select any positions yet.)

	We show that the loop invariant is maintained.
	Assume that the loop invariant is true up to iteration $l$.
	Consider first a loop iteration where we do not return in lines 6 or 10.
    As such we do not take the branch in line 3, and
	$\textsc{SelectCutPosition}$ selects a position $i\in K(\tilde{P}_{l})$.

	We construct two paths $P_1,P_2$ in line 8 by cutting $\tilde{P}_{l}$.
	Because of the application of \Cref{lem:merging_paths2} in line 9, $\tilde{x}_{l+1}$ differs from $\tilde{x}_l$ only for variables in $X(P_1)\cup X(P_2)$ (by the statement of that lemma). By \Cref{lem:alternation_property2}, all variables
	in $X(P_1)$ or all variables in $X(P_2)$ become integral.
    Therefore, $\tilde{P}_{l+1}$ is stilla critical path after the assignment in line 11.
	\Cref{lem:merging_paths2} ensures that $\tilde{x}_{l+1}$ remains valid for $LP_{\mathcal{I},\tilde{\mathcal{F}}_{l}\cup I'}$, where $I'$ consists of exactly the start and end positions of $P_1$ and $P_2$.
    However, the start position of $P_1$ and the end position of $P_2$ are exactly the start position
	of $\tilde{P}_{l}$ and are already in $I^C_{l}$. The end position of $P_1$ and the start position of $P_2$ are identical, and are the position $i$ selected in line 7. By definition, $\tilde{\mathcal{F}}_{l+1}=\tilde{\mathcal{F}}_{l}\cup \set{i}$. Therefore, $\tilde{x}_{l+1}$ is valid for $LP_{\mathcal{I},\tilde{\mathcal{F}}_{l+1}}$.
    $\tilde{P}_{l+1}$ then gets assigned the remaining fractional path in line 11 (if both paths became integral, then we would have returned in line 10).
    
    Furthermore, $\tilde{P}_{l+1}$ is still a critical path:
    $\tilde{P}_{l}$ was critical in $\tilde{x}$, so the fractional variables are exactly the ones in $X(\tilde{P}_l)=X(P_1)\cup X(P_2)$. One of the paths became integral, and $\tilde{P}_{l+1}$ was assigned the other fractional path in line 11. Therefore
    $X(\tilde{P}_{l+1})$ now contains exactly the fractional variables in $\tilde{x}_{l+1}$, making it a critical path.
    By \Cref{lem:alternation_property2}, $\tilde{P}_{l+1}$ is then alternating.
    By assumption, the start and end position of $\tilde{P}_{l}$ are in $I^C_l\subseteq I^C_{l+1}$, so after
    adding the position $i$ selected in line 7 to $I^C_{l+1}$, the start and end positions of $\tilde{P}_{l+1}$
    are still in $I^C_{l+1}$.

    Now consider the three lines where \textsc{RepeatedCut} can return a solution (lines 6, 10 and 13).
    First consider the case that \textsc{RepeatedCut} returns in line 10 in the $l$'th iteration.
    The loop invariant held at the beginning of the iteration and $K(\tilde{P}_{l})\ne\varnothing$.
    We applied \Cref{lem:merging_paths2} in line 9. Because both paths $P_1,P_2$ constructed in line 8
    became integral and $\tilde{P}_l$ was critical, the solution $\tilde{x}_*$ that gets returned is integral.
    Furthermore, $\tilde{x}_*$ is a solution for $LP_{\mathcal{I},\tilde{\mathcal{F}}_*}$ by the same arguments
    as above.
    A similar argument holds when the algorithm returns in line 13. The only difference is that $\tilde{P}_l$ is sufficient to apply \Cref{lem:merging_paths2} since $\Delta(\tilde{P}_l)=0$.
    
    Lastly, suppose that we return in line 6.
    Since $K(\tilde{P}_l)=\varnothing$, and $\tilde{P}_l$ is critical by the validity of the loop invariant,
    there are only two fractional variables (namely the ones in $X(\tilde{P}_l)$) in $\tilde{x}$.
    By rescheduling the job $j\in J(\tilde{P}_l)$ in line 5, the resulting solution $\tilde{x}_{*}$ becomes integral.
    $\tilde{x}_{l}$ is a valid solution for $LP_{\mathcal{I},\tilde{\mathcal{F}}_l}$ by the loop invariant, where
    $I^C_l$ contains the start and end position of $\tilde{P}$.
    $j$ is rescheduled into one of these two positions, so the returned solution $\tilde{x}_*$ is valid for $LP_{\mathcal{I},\tilde{\mathcal{F}}_l}=LP_{\mathcal{I},\tilde{\mathcal{F}}_*}$.
\end{proof}

\algrepeatedcutproperties*
\begin{proof}
    The cost $C(\tilde{x})$ is initially equal to $C(x)$ as we set $\tilde{x}\gets x$ in line 1. Afterwards, $\tilde{x}$ may only be changed in lines 5,9 and 12.
    For lines 9 and 12, \Cref{lem:merging_paths2} guarantees that the cost of $\tilde{x}$ does not increase.

    Now consider the change in line 5 and assume that it happens at the $l$'th iteration of the algorithm.
    Since $\tilde{P}_l$ is alternating (see \Cref{lem:alg_offline:loop_invariant}), we have
    $x_j^O=x_{j,i_1,t_{j^O}}=y$ and $x_j^E=x_{j,i_2,t_j^E}=1-y$ for some $y\in\intoo{0,1}$.
    First consider the case that $j\in J(\tilde{\mathcal{F}}_*)$.
    Then $t_j^O=t_j^E$, and therefore $j$ will be rescheduled into position $\min(i_1,i_2)$ in line 5 of the algorithm of type $t_j^O$.
    Its contribution will change from $i_1\cdot p_j^{t_j^O}\cdot y+i_2\cdot p_j^{t_j^O}\cdot (1-y)$ to
    $\min (i_1,i_2)\cdot  p_j^{t_j^O}$, which decreases the objective, so
    $C(\tilde{x})\le C(x)$.
    Otherwise, rescheduling $j$ will at most add its new contribution to the cost, giving
    $C(\tilde{x})\le C(x)+Z$ where $Z$ is $j$'s contribution.
\end{proof}

\algofflinemovingjobs*
\begin{proof}
    We calculate the contribution for a job $j\in J$:
    \begin{align*}
        \sum_{i\in I,t\in T} i \cdot p_j^t \cdot \tilde{x}_{j,i,t}
        =i' \cdot p_j^{t_j}
        =\frac{i'}{i} \cdot \left(i \cdot p_j^{t_j}\right)=\frac{i'}{i} \sum_{i\in I,t\in T} i \cdot p_j^t \cdot x_{j,i,t},
    \end{align*}
    so the contribution of job $j$ increased by a factor of $i'/i$.
\end{proof}

\mufunctionupdate*
\begin{proof}
    We will gradually update the charging function with every iteration of \textsc{RepeatedCut}.
    We first observe
    the $l$'th iteration, specifically how the charging set $I^+_{l+1}$ compares to $I^+_{l}$.
	First, $\textsc{SelectCutPosition}$ selects a position $i\in K(\tilde{P}_l)$ that is among
    the smallest $2/\epsilon+1$ positions in $K(\tilde{P}_l)$ (or all of them, if there are less than $2/\epsilon+1$ many).
    Recall that $I^+_{\bar{l}}$ contains all positions from $I^C_{\bar{l}}$ for each iteration $\bar{l}$, as well as a set of positions $I'$.
    Here, $I'$ consists of the smallest $2/\epsilon+1$ positions in $K(\tilde{P}_l)$ (or all of them, if there are less), as computed in line 2 of \textsc{SelectCutPosition}.
    By definition, $I^C_{l+1}=I^C_{l}\cup\set{i}$ and $i\in I^+_l$.
    
    In the $l$'th iteration,
	the variables of at least one of the paths $P_1,P_2$ constructed in line 8 of \textsc{RepeatedCut} become integral, since
	\Cref{lem:alg_offline:loop_invariant} guarantees that $\tilde{P}$ remains critical after the assignment in
    line 11 of $\textsc{RepeatedCut}$ when we transition to the $(l+1)$'st iteration.
    
    First assume that $\abs{K(\tilde{P}_l)}\ge 2/\epsilon+1$.
    Because we return the position that appears as $(1/\epsilon+1)$'st position in $\tilde{P}$ of the positions
    in $I'$, $P_1$ and $P_2$ will each contain exactly $1/\epsilon$ many positions of $I'$.
    Since \textsc{SelectCutPosition} only gets to select positions from $K(\tilde{P})$, the positions of $K(P_1)$ or
    the positions of $K(P_2)$ will not be in $I^+_{l+1}$.
    Because the smallest $2/\epsilon+1$ positions are considered by $\textsc{SelectCutPosition}$, there are always $1/\epsilon$ of these positions that will be considered by the next iteration of $\textsc{SelectCutPosition}$.

    Summarizing, we can split $I'$ into three sets:
    A set of $1/\epsilon$ positions that will be in $K(\tilde{P}_{l+1})$,
    a set $I_1$ of $1/\epsilon$ positions that were in $K(P')$ where $P'$ was the path that became integral in line 9 of \textsc{RepeatedCut},
    and one position $i$ that
    was selected by $\textsc{SelectCutPosition}$ in the $l$'th iteration.
    We already mentioned that $i\in I^+_l$, so
    $I^{+}_{l+1}=(I^{+}_l\setminus I_1)\cup I_2$, where $I_2$ contains the next-largest
    $1/\epsilon+1$ positions from $K(\tilde{P}_{l+1})$ (or all positions from $K(\tilde{P}_{l+1})$, if $\abs{K(\tilde{P}_{l+1})}<1/\epsilon+1$).

    We describe how to update the charging function.
    We let $\mu_{l+1}\coloneqq \mu_l$ and then update $\mu_{l+1}$ as follows:
    Take any subset $I_3\subseteq I_2$ with $\abs{I_3}=1/\epsilon$, and take any bijection
    $b: I_3\rightarrow I_1$. Then we update $\mu_{l+1}(b(\bar{i}))\gets\mu_{l}(\bar{i})$ for all $\bar{i}\in I_3$. This is justified since
    all positions in $I_2$ are larger than the positions in $I_1$ by definition.
    It remains to assign one last position $\bar{i}\in I_2\setminus I_3$.
    We set $\mu_{l+1}(\bar{i})\gets I_1$.
    This is also valid since $I_1\cap I^{+}_{l+1}=\varnothing$ and all positions in $I_1$ are smaller than $\bar{i}$ (again since all positions in $I_2$, especially $\bar{i}$, are larger than the positions in $I_1$).

    If we assume that $\abs{K(\tilde{P}_l)}< 2/\epsilon+1$, then $I^+_{l}$ already contains all of these
    positions. Therefore $I^+_{l+1}\subseteq I^+_l$ must hold, and we can find $\mu_{l+1}$ by restricting
    $\mu_l$ to $I^+_{l+1}$.
    By this argument, it is easy to find a charging function $\mu_*$.
\end{proof}

\findilpsolution*
\begin{proof}
    By \Cref{lem:alg_repeated_cut_properties}, $x$ is an integral valid solution
    for $LP_{\mathcal{I},\mathcal{F}}$, where $I^C(F)\subseteq I^+_l$.
    Since $x$ is integral, each position in $I\setminus I^C(F)$ schedules exactly one job, and each position in $I^C(F)$ schedules $0$, $1$ or $2$ jobs.
    
    We reschedule the jobs to new positions as follows:
    Scan through the positions $1,\dots,n$ in ascending order.
    If two jobs are scheduled in the current position $i$,
    look for the smallest position $i'>i$ where no job is scheduled.
    (This position must exist as there is exactly one position per job.)
    Now move all jobs from positions $\bar{i}\in\set{i+1,\dots,i'-1}$ to the
    respective position $\bar{i}+1$. Additionally, move one job from position $i$ to position $i+1$.
    Now all positions up to position $i+1$ schedule exactly one job.
    Repeat this procedure until each position schedules exactly one job, so we
    end up with a valid solution $\tilde{x}$ for $ILP_{\mathcal{I}}$.

    With this we have moved each job to a new position, and can analyze the increase in its
    contribution with \Cref{obs:alg_offline_moving_jobs}.
    Consider any job $j$. Let $i$ be its position in $x$, and $i'$ be its position in $\tilde{x}$.
    By \Cref{obs:alg_offline_moving_jobs}, we have to show that $i'/i\le 1+\epsilon$.
    
    By the above operation, $j$'s position is incremented by one at most $k$ times, where
    $k$ is the number of positions in $I^+_*$ that are no larger than $i$.
    It follows that $i'\le i+k\le i+\abs{\set{\bar{i}\in I^+_* | \bar{i}\le i}}$.
    We can now make use of the charging function $\mu_*$.
    For each position $\bar{i}\in I^+_*$, there exist $1/\epsilon$ positions $\mu_*(\bar{i})$ that are not in $I^+_*$.
    Furthermore, all these positions are smaller than $i$, and the positions that $\mu_*$ maps
    to are disjoint for any two positions in $I^+_*$.
    It follows that
    $i\le \abs{\set{\bar{i}\in I^+_* | \bar{i}\le i}}\cdot (1+1/\epsilon)$.
    With this, we can bound the ratio
    \begin{align*}
        \frac{i'}{i}
        \le \frac{i+\abs{\set{\bar{i}\in I^+_* | \bar{i}\le i}}}{i}
        \le \frac{i+i/(1+1/\epsilon)}{i}
        = 1+\frac{\epsilon}{1+\epsilon}
        \le 1+\epsilon
    \end{align*}
    \vspace{-3em}\\\vspace{0.5em}
\end{proof}

\initialsupportfunction*
\begin{proof}
    By definition, $I^+_1$ consists of the smallest $2/\epsilon+1$ positions in the critical path $\tilde{P}_1$. The smallest $M=(2/\epsilon+1)/\epsilon$ positions can not lie in $I(\tilde{P}_1)$ since $\tilde{P}_1$ is a fractional path (see \Cref{lem:alg_offline:loop_invariant}). Then we can find $\mu_1$ by assigning each position in $I^+_1$ a distinct set of $1/\epsilon$ of the $M$ positions.
    This shows the existence of a charging function $\mu_1$ for $I^+_1$.
\end{proof}

\offlineptas*
\begin{proof}%
	Let $\mathcal{I}$ be an instance, and $x^*$ be an optimal solution to $ILP_{\mathcal{I}}$.
	Then there exists a fixation $\mathcal{F}^*$ (defined in \Cref{def:fixation_brute_force}) that matches $x^*$, i.e., such that $x^*$ is a solution to $LP_{\mathcal{I},\mathcal{F}^*}$.
	We first brute-force $\mathcal{F}^*$.
	Specifically, we have to brute-force $X(\mathcal{F}^*)$, fixing the jobs that are scheduled in positions $1,\dots,M$ and their type.
	For each of the $M$ positions, there are $2n$ choices ($n$ for the job count and $2$ choices for the type of the job).
	In total, there are $(2n)^M=2n^{(2/\epsilon+1)/\epsilon}$ possibilities to consider, which is polynomial in $n$.

    We calculate an optimal solution $x$ to $LP_{\mathcal{I},\mathcal{F}^*}$.
	Then $C(x)\le C(x^*)$.
	We apply \Cref{lem:find_decomposition2} to $x$, obtaining a solution $x'$ for $LP_{\mathcal{I},\mathcal{F}^*}$ with $C(x')\le C(x)$.
	If $x'$ is already integral, then we are done, as then $x'$ is a solution to $ILP_{\mathcal{I}}$ and its cost is no larger than the cost of $x^*$.
	Otherwise, \Cref{lem:find_decomposition2} gives us a blocking cycle $P$ for $x'$.
 
	We apply $\textsc{RepeatedCut}$ for a solution $x'$ for $L$ and blocking cycle $P$ for $x'$.
	(Note that $I^C(\mathcal{F}^*)=\varnothing$, so that algorithm is applicable.)
	$\textsc{RepeatedCut}$ then returns a solution $\tilde{x}$.

    We apply \Cref{lem:alg_repeated_cut_properties}.
    Suppose that it gives us $C(\tilde{x})\le C(x')+Z$ where
    $Z$ is the contribution of a job $j$ that does not lie in $J(\mathcal{F}^*)$.
    Then we can apply \Cref{lem:find_ilp_solution} (using the charging function provided by \Cref{obs:initial_support_function}) to obtain a solution where the contribution of each job
    is increased by a factor of at most $1+\epsilon$.
    By removing $j$, we obtain an (invalid) solution $\tilde{x}'$ with $C(\tilde{x}')\le (1+\epsilon)\cdot C(x')$.
    We will then schedule $j$ of type $\wedge$ at position $M+1$, moving all jobs
    from positions $i\ge M+1$ to the respective position $i+1$.
    Doing this produces a schedule $x''$ that is valid for $ILP_{\mathcal{I}}(x'')$.
    
    Since $j\notin J(\mathcal{F}^*)$, we must have $p_j^\wedge\le \min_{i\in [M]}{p_{j_i}^{t_i}}$, where $j_i$ is the job (brute-force-) scheduled in position $i$ of type $t_i$.
    Therefore, inserting $j$ at position $M+1$ adds a cost of
    \begin{align*}
        (M+1) \cdot p_j^\wedge
        \le \frac{2}{M} \cdot \frac{M(M+1)}{2} \cdot p_j^\wedge
        =\frac{2}{M}\cdot p_j^\wedge\cdot \sum_{i=1}^{M} i
        \le \frac{2}{M} \sum_{i=1}^{M} i\cdot p_{j_i}^{t_i}
        \le \frac{2}{M} C(x^*)
    \end{align*}
    We then can make use of \Cref{obs:alg_offline_moving_jobs}.
    The contribution of each job thus increases by a factor of at most $(M+1)/M$.
    We conclude that
    \begin{align*}
        C(x'')
        &\le \frac{M+1}{M} (C(\tilde{x}')+(M+1) \cdot p_j^\wedge)
        \le \frac{M+1}{M} \left((1+\epsilon)\cdot C(x')+\frac{2}{M}\cdot C(x^*)\right)\\
        &\le \frac{M+1}{M} \left((1+\epsilon) +\frac{2}{M}\right)\cdot C(x^*)
        =(1+\LDAUOmicron{\epsilon})\cdot C(x^*)
    \end{align*}

    It is easy to see that if \Cref{lem:alg_repeated_cut_properties}
    yields a solution $\tilde{x}$ with $C(\tilde{x})\le C(x')$ that we can also bound the cost of the resulting integral solution in a similar way. The only difference is that there is no job $j$ that we need to insert.
\end{proof}

\section{Hardness of \SLPBtc/}
\label{sec:hardness}

In this section, we prove  \Cref{thm:l1norm:offline:hardness}.
Before stating the formal proof, we give the high level-idea of the reduction in the following:

\noindent\textbf{Reduction Overview.}
The NP-hardness (formally given in \cref{sec:hardness}) is via a reduction from the \textsc{Partition} problem.
For a given \textsc{Partition} instance with positive integers $U = \set{u_1, \ldots, u_n}$, we construct an instance $\cI = (J, \vph, \vpc=\0, \vc, B)$ of \SLPBtc/ such that the minimal total completion time reaches a certain value if and only if there is an $S \subseteq U$ whose elements sum up to $U_{\mathrm{half}} \coloneqq \sum_{j=1}^{n} u_j/2$.
More exactly, for each element $u_j \in U$ we create two jobs $j_1, j_2$ such that for their testing cost and (upper) processing times we have $c_{j_1} -c_{j_2} = u_j$ and $\ph_{j_1}-\ph_{j_2}=u_j/j$.
This means that testing $j_1$ is more expensive than testing $j_2$ but yields also a higher reduction of the total completion time.
If we test all cheap jobs $\set{j_2 | j \in [n]}$, the total testing cost is $\sum_{j=1}^{n} c_{j_2}$.
If we pick all expensive jobs $\set{j_1 | j\in [n]}$, we require a budget of $\sum_{j=1}^{n} c_{j_2}+2U_{\mathrm{half}}$.
We set the budget to $B = \sum_{j=1}^{n} c_{j_2} + U_{\mathrm{half}}$ to force an optimal solution for \SLPBtc/ to select jobs with total testing cost of exactly $B$ and thus indirectly find a partition.

Now consider a partition $(S,U\setminus S)$, assuming $\sum_{u\in S}u \leq U_{\mathrm{half}}$.
Such a partition corresponds to a feasible testing plan for instance $\cI$: for each $j \in \intcc{n}$, test job $j_1$ if $u_j \in S$ and job $j_2$ otherwise.
We can construct the job set such that $\forall j' < j$: $\ph_{j'_2}> \ph_{j_2}$ and for each job pair $j_1,j_2$, either they are neighboring in the optimal schedule or at least one of them is tested.
Then we can show that the aforementioned testing plan obtains a total completion time of $\sum_{j=1}^{n}j\cdot \ph_{j_2} + \sum_{u_j\in U\setminus S} u_j$. Thus, one direction of the reduction can be proved: if the partition instance $U$ is a yes-instance, then
the minimum total completion time of instance $\cI$ is at most $\sum_{j=1}^{n}j\cdot \ph_{j_2} + U_{\mathrm{half}}$.

For the other direction's proof, 
an observation is that for the testing plans that test exactly one job in $\{j_1, j_2\}$ for each $j\in [n]$, there exists such a plan with a total completion time $\sum_{j=1}^{n}j\cdot \ph_{j_2} + U_{\mathrm{half}}$ only if $U$ is a yes-instance.
We finally complete the proof by showing that we can find appropriate $\{\ph_{j_2}, c_{j_2}\}_{j\in [n]}$ such that once a solution does not test exactly one job in $\{{j_1}, {j_2}\}$ for each $j\in [n]$, either it violates the budget constraint or its objective value is larger than $\sum_{j=1}^{n}j\cdot \ph_{j_2} + U_{\mathrm{half}}$.

\begin{theorem}
    \label{thm:l1norm:offline:hardness}
    The problem \SLPB/ is $\NP$-hard, even for instances $\mathcal{I}=(J,\vph,\vpc,\vc,B)$ where $\vpc=\mathbf{0}$.
\end{theorem}

\begin{proof}
    We reduce from the problem \textsc{Partition}, where we are given a set $U=\set{u_1,\dots,u_n}$ of positive integers to partition.
    We may assume that $U$ is a yes-instance. Then it is NP-hard to find a partitioning
    of $U$.
    
    Because $U$ is a yes-instance, we have $u_j\le 1/2\cdot \sum_{i=1}^n u_i$ for all $j\in [n]$.
    For each $u_j\in U$, generate two jobs $j_1$ and $j_2$.
    We first (recursively) define values $A_1,\dots,A_n$ and $B_1,\dots,B_n$ as follows, and then give the upper limit processing times and the probing costs of the two jobs.
    \begin{align*}
        A_j= \frac{1}{j+1}\left(\sum_{i=j+1}^n {i A_i}+\frac12\sum_{i=1}^n u_i+1-u_j\right)~~~~~~~~~~~~~~~~~~
        B_j= \left(\sum_{i=j+1}^{n}{B_i}\right)+\frac12\left(\sum_{i=1}^{n}{u_i}\right)+1
    \end{align*}
    Note that $A_j>0$ for all $j\in [n]$ since $u_j\le 1/2\cdot \sum_{i=1}^n u_i$.
    The $B_j$-values are trivially positive.
    Based on these values, we define the job parameters as
    \begin{align*}
        \ph_{j_1}=A_j+u_j/j~~~&~~~\ph_{j_2}=A_j~~~&~~~c_{j_1}=B_j+u_j~~~&~~~c_{j_2}=B_j
    \end{align*}
    Furthermore, we set the budget to be $B=\sum_{j=1}^{n} B_j+1/2\cdot \sum_{j=1}^n u_j$.
    Lastly, we set all tested processing times to zero.
    This finishes the construction of an instance $\mathcal{I}=(J,\vph,\mathbf{0},\vc, B)$.

    Consider any set $\bar{U}\subseteq U$, and let the \emph{corresponding set of tested jobs} be
    $J_{\vee}=\set{j_1 | u_j\in \bar{U}}\cup \set{j_2 | u_j\in U\setminus \bar{U}}$.
    Similarly, if we have a set $J_{\vee}$ where from each pair of jobs $\{j_1,j_2\}$ exactly one job is tested, we can find a corresponding set $\bar{U}$.
    
    For such a set of tested jobs, we will show that $S=(J_{\vee},\sigma)$ with optimally chosen $\sigma$ we have
    \begin{equation}
        \label{thm:l1norm:offline:hardness:sum_equivalence}
        \sum_{k\in J_{\vee}} c_{k}+C(S)=\sum_{j=1}^{n} \left(B_j+j\cdot A_j+u_j\right)
    \end{equation}
    To see this, we calculate the total budget used ($\sum_{k\in J_{\vee}} c_k$) and the total completion time $C(S)$ of an optimal schedule that tests $J_{\vee}$.
    
    The total budget used is
    \begin{equation}
        \label{thm:l1norm:offline:hardness:budget_use}
        \sum_{u_j\in \bar{U}}{c_{j_1}}+\sum_{u_j\in U\setminus \bar{U}}{c_{j_2}}
        =\left(\sum_{u_j\in \bar{U}}{B_j+u_j}\right)+\sum_{u_j\in U\setminus \bar{U}}{B_j}
        =\left(\sum_{j=1}^{n} B_j\right)+\sum_{u_j\in \bar{U}}{u_j}
    \end{equation}

    We also get that $\ph_{j'_1}>\ph_{j'_2}>\ph_{j_1}>\ph_{j_2}$ for any $j'<j$:
    The first and the last inequality are trivial.
    The second one evaluates to
    \begin{align*}
        A_{j-1}>A_j+\frac{u_j}{j}\Leftarrow
        \frac{1}{j}\left(\sum_{i=j}^n {i A_i}+\frac12\sum_{i=1}^n u_i+1-u_{j-1}\right)>A_j+\frac{u_j}{j}\Leftarrow
        \frac{1}{j}\left(\frac12\sum_{i=1}^n u_i+1-u_{j-1}\right)>\frac{u_j}{j}
    \end{align*}
    where the last inequality follows from $u_j\le 1/2\cdot \sum_{i=1}^n u_i$ for all $j\in [n]$.

    Since $J_{\vee}$ tests exactly one job $\set{j_1,j_2}$ for each $j\in [n]$, $\sigma$ will schedule the untested job at position
    $2n-j+1$ for each $j$ and all tested jobs arbitrarily in the first $n$ positions.

    Only the untested jobs contribute to the total completion time realized by $S$.
    The total completion time can then be calculated as
    \begin{align*}
        C(S) = \sum_{j=1}^{n} A_j \cdot j+\sum_{j_1\notin J_{\vee}}{u_j}=\sum_{j=1}^n A_j \cdot j+\sum_{u_j\in U\setminus \bar{U}}{u_j}
    \end{align*}

    Summing both, we get
    \begin{align*}
        \sum_{k\in J_{\vee}} c_k+C(S)=
        \sum_{j=1}^{n} B_j+\sum_{u_j\in \bar{U}}{u_j}+\sum_{j=1}^n A_j \cdot j+\sum_{u_j\in U\setminus \bar{U}}{u_j}=\sum_{j=1}^{n} \left(B_j+j\cdot A_j+u_j \right)
    \end{align*}

    Now assume that there exists some $U^*\subseteq U$ which partitions $U$.
    Then the budget used (according to \Cref{thm:l1norm:offline:hardness:budget_use}) is
    \begin{align*}
        \left(\sum_{j=1}^{n} B_j\right)+\sum_{u_j\in U^*}{u_j}=\left(\sum_{j=1}^{n} B_j\right)+\frac12\cdot \sum_{j=1}^n{u_j}=B
    \end{align*}
    Because of \Cref{thm:l1norm:offline:hardness:sum_equivalence}, the corresponding set of tested jobs $J_{\vee}$ of $U^*$ must minimize the total completion time of the schedule $S=(J_\vee,\sigma)$ with optimally chosen $\sigma$ among all $\bar{U}\subseteq U$. The total completion time is
    \begin{equation}
        \label{thm:l1norm:offline:hardness:completion_time}
        C(S)=\sum_{j=1}^{n} B_j+j\cdot A_j+u_j-\sum_{k\in J_{\vee}} c_k=\sum_{j=1}^{n} B_j+j\cdot A_j+u_j-B=
        \sum_{j=1}^{n} j\cdot A_j+\frac12 u_j
    \end{equation}

    It remains to show that exactly one job of each pair must be tested in an optimal valid schedule $S^*$.
    Suppose the contrary and let $j$ be smallest such that either both $j_1,j_2$ are tested, or both are not tested.
    Suppose the case that both jobs are tested.
    Clearly, the budget used by $S^*$ is at least
    \begin{align*}
        \left(\sum_{i=1}^{j-1} c_{i_2}\right) + c_{j_1}+c_{j_2}
        &=\left(\sum_{i=1}^{j-1} B_i\right) + 2 B_j+u_j
        \ge \left(\sum_{i=1}^{j-1} B_i\right) + B_j+\left(\sum_{i=j+1}^{n}{B_i}\right)+\frac12\left(\sum_{i=1}^{n}{u_i}\right)+1\\
        &=\left(\sum_{i=1}^{n} B_i\right)+\frac12\sum_{i=1}^{n}{u_i}+1>B
    \end{align*}
    with a contradiction to the validity of $S^*$.

    Instead, assume that neither of the jobs is tested.
    Since $\ph_{j_2} < \ph_{j_1}$ for all $j\in [n]$, the total completion time of $S^*$ is
    \begin{align*}
        C(S^*)\ge& \left(\sum_{i=1}^{j-1} i \cdot \ph_{i_2}\right)+j \cdot \ph_{j_1}+(j+1)\cdot \ph_{j_2}
        =\left(\sum_{i=1}^{j-1} i \cdot A_i\right)+j \left(A_j+\frac{u_j}{j}\right)+(j+1) A_j\\
        =&\left(\sum_{i=1}^{j} i \cdot A_i\right)+u_j+(j+1) A_j
        =\left(\sum_{i=1}^{n} i \cdot A_i\right)+\left(\frac12\sum_{i=1}^n u_i+1\right)
    \end{align*}
    This is larger than the total completion time of the schedule in \Cref{thm:l1norm:offline:hardness:completion_time}, a contradiction.
    
    It is easy to see that $\mathcal{I}$ is polynomial in the input size.
    This finishes the reduction.
\end{proof}

\section{Oblivious \SLPBtc/}
\label{sec:oblivious}
In this section, we consider the oblivious version of the problem under the objective of total completion time minimization (\SLPBtc/).
Recall that the lower processing time vector $\vpc$ is the only hidden information for the algorithm. 
We first give a lower bound of the oblivious problem and then introduce the algorithmic framework.

\begin{restatable}{theorem}{onlinecompletehardness}
\label{thm:online:l1norm:hardness}
For oblivious \SLPBtc/, there is no deterministic algorithm whose competitive ratio is strictly smaller than $4$ even for the setting of uniform testing cost.
\end{restatable}
\begin{proof}
We prove the theorem by giving the following hard instance.
The instance consists of a job set $J:=[n]$ with $n$ jobs, and the required testing budget for each job is $1$, i.e., $c_j=1$ for all $j\in[n]$.
All jobs have the same upper limits on the processing time $1$, i.e., $\vph=\mathbf{1}$.
Let $B:=\frac{n}{2}$ be the total amount of the budget.
Note that any solution can test at most $\frac{n}{2}$ jobs.
Let $\alg$ be an arbitrary deterministic algorithm.
Let $S$ be a set of jobs tested by $\alg$.
The adversary can always make the testing operation of $\alg$ useless.
Namely, the adversary can set up an instance such that (\rom{1}) $\pc_j=1$ for all $j\in S$; (\rom{2}) $\pc_j=0$ for all $j\in J\setminus S$.
Thus, for any algorithm $\alg$, there always exists an instance $\cI$ such that $\alg(\cI)=1+2+\cdots+n=\frac{n(n+1)}{2}$ while the optimal solution $\opt(\cI)=1+2+\cdots+\frac{n}{2}=\frac{n(n+2)}{8}$.
Therefore, any deterministic algorithm has a competitive ratio of at least $4$.
\end{proof}

Now, we present a $(4+\epsilon)$-competitive algorithm (\cref{alg:online:l1norm}) and show the following main result (\cref{thm:online:l1norm:ratio}).
Note that the competitive ratio is essentially optimal by the hardness result we have shown in \cref{thm:online:l1norm:hardness}.

\begin{theorem}
    Given an arbitrary instance $\cI:=(J,\vph,\vpc,\vc,B)$ with $\vpc$ is hidden for the algorithm, \cref{alg:online:l1norm} is a $(4+\epsilon)$-competitive algorithm.
    Moreover, \cref{alg:online:l1norm} is a $4$-competitive algorithm when $\vc=\mathbf{1}$.
    \label{thm:online:l1norm:ratio}
\end{theorem}

In the following, we first give some intuitions of the algorithm and then present the formal description of the complete algorithm and the analysis later.

\paragraph{Algorithmic Framework} 
Our algorithm is inspired by the hard instance described in the proof of \cref{thm:online:l1norm:hardness}.
Intuitively, the adversary can always make the testing of an algorithm useless, which is the worst case scenario for the algorithm. 
In contrast, the testing made by the optimal solution can reduce the most objective value.
That is, the lower processing time of the jobs tested by the optimal solution becomes $0$ while the lower processing time of the jobs tested by the algorithm is the same as their upper bound.
Knowing such a property of the optimal solution, our algorithm will try to simulate the "behavior" of the optimal solution.
That is, we can pretend that the processing time of a job became $0$ when it was tested by the algorithm.
This is actually equivalent to solving an offline instance in which $\vpc=\mathbf{0}$ and all other parameters are the same as the oblivious instance.
For convenience, we define such a constructed instance as the {\em auxiliary instance}.
See \cref{def:aux-instance} for the formal definition.

\begin{definition}[Auxiliary Instance]
Given an arbitrary instance $\cI:=(J,\vph,\vpc,\vc,B)$, let $\wtI:=(\widetilde{J},\widetilde{\vph},\widetilde{\vpc},\widetilde{\vc},\widetilde{B})$ be the {\em auxiliary instance} of $I$, where each entry of $\wtI$ is defined as follows: 
$\widetilde{J}\leftarrow J$;
$\widetilde{\vph}\leftarrow \vph$;
$\widetilde{\vpc}\leftarrow \mathbf{0}$;
$\widetilde{\vc}\leftarrow \vc$;
$\widetilde{B}\leftarrow B$.
For notation convenience, we also write $\wtI:=(J,\vph,\mathbf{0},\vc,B)$ as the auxiliary instance of $\cI$.
\label{def:aux-instance}
\end{definition}

\subsection{The Complete Algorithm}

We now present the complete algorithm (\cref{alg:online:l1norm}) by implementing the algorithmic framework discussed in the previous section.

\begin{algorithm}[htb]
    \caption{\textsc{The Complete Algorithm}}
    \label{alg:online:l1norm}
    \begin{algorithmic}[1]
        \REQUIRE An instance $\cI:=(J,\vph,\vpc,\vc,B)$ where $\vpc$ is hidden for the algorithm.
        \ENSURE A set $S\subseteq J$ of tested jobs.
        \STATE Construct the auxiliary instance $\wtI:=(J,\vph,\mathbf{0},\vc,B)$.
        \STATE Solve the constructed auxiliary instance $\widetilde{\cI}$ and let $S$ be the returned solution.
        \RETURN $S$.
    \end{algorithmic}
\end{algorithm}

The analysis shows that \cref{alg:online:l1norm} computes a $(4+\epsilon)$-approximation solution to \SLPBtc/ in polynomial time consisting of the following two steps.

\begin{enumerate}
    \item Given an arbitrary oblivious instance $\cI$, we show that there is an FPTAS for the auxiliary instance $\wtI$ of $\cI$. 
    Formally, we prove the following lemma in \cref{subsec:online:aux}.
    \begin{restatable}{lemma}{auxinstance}
    Given an arbitrary instance $\cI:=(J,\vph,\vpc,\vc,B)$ with $\vpc=\mathbf{0}$, there is a pseudo-polynomial time algorithm that returns an optimal solution. 
    Moreover, such a pseudo-polynomial time algorithm can be converted into an FPTAS. 
    \label{lem:online:l1norm:pc=0}
    \end{restatable}

    Note that we have a PTAS for the general instance by \cref{thm:offline:ptas}, but we can get an FPTAS for the auxiliary instance, which is a special case of the general instance.
    As a corollary, the same algorithm can achieve a better approximation when $\vc=\mathbf{1}$.
    
    \begin{corollary}
    Given an arbitrary instance $\cI:=(J,\vph,\vpc,\vc,B)$ with $\vpc=\mathbf{0}$ and $\vc=\mathbf{1}$, there is a polynomial time algorithm that returns an optimal solution.
    \label{cor:online:l1norm:pc=0+c=1}
    \end{corollary}
    \item We show that the worst-case instance of \cref{alg:online:l1norm} must satisfy the following two assumptions: (\rom{1}) the optimal solution tests exactly the jobs that \cref{alg:online:l1norm} does not test; (\rom{2}) all jobs $j$ that are tested by the optimal solution have $\pc_j=0$ and all jobs $j$ that are tested by \cref{alg:online:l1norm} have $\pc_j=\ph_j$.
    Formally, we prove the following lemma in \cref{subsec:online:worst-instance}.
    \begin{restatable}{lemma}{worstinstance}
    Given an arbitrary instance $\cI:=(J,\vph,\vpc,\vc,B)$, let $\alg(\cI)$ and $\opt(\cI)$ be the set of tested jobs chosen by \cref{alg:online:l1norm} and optimal solution, respectively.
    Let $\com(\alg(\cI))$ and $\com(\opt(\cI))$ be the objective value of the algorithm's and the optimal solution, respectively.
    There must exist another instance $\cI'$ such that the following properties are true:
    \begin{enumerate}[label=(P\arabic*),leftmargin=*,align=left]
        \item $\frac{\com(\alg(\cI))}{\com(\opt(\cI))} \leq \frac{\com(\alg(\cI'))}{\com(\opt(\cI'))}$;
        \label{online:worst-instance:pro:1}
        \item $\alg(\cI')\cap\opt(\cI')=\emptyset$;
        \label{online:worst-instance:pro:2}
        \item $\alg(\cI')\cup\opt(\cI')=J$;
        \label{online:worst-instance:pro:3}
        \item $\pc_j=\ph_j$ for all $j\in \alg(\cI')$ and $\pc_j=0$ for all $j\in \opt(\cI')$.
        \label{online:worst-instance:pro:4}
    \end{enumerate}
    \label{lem:online:l1norm:worst-intance}
    \end{restatable}
\end{enumerate}

Given the above two key lemmas, we prove that \cref{alg:online:l1norm} is a $(4+\epsilon)$-approximation algorithm for the oblivious setting in the following.

\begin{proofof}{\cref{thm:online:l1norm:ratio}}
Given an arbitrary instance $I:=(J,\vph,\vpc,\vc,B)$, let $\alg(I)$ and $\opt(I)$ be the set of tested jobs chosen by \cref{alg:online:l1norm} and the optimal solution, respectively.
Without loss of generality, we assume that the instance $I$ is the worst-case instance and thus satisfies the properties stated in \cref{lem:online:l1norm:worst-intance}.
We use $\alg$ and $\opt$ to denote the value of the algorithm's and optimal solution, respectively.
In this proof, we only care about the upper limit of the processing time of each job.
Therefore, for each job $j$, we refer $p^{(j)}$ as its $\ph_j$.

We assume the optimal solution has exactly $m$ untested jobs, i.e., $\abs{J\setminus\opt(I)}=m$.
Let $p^{(1)}\geq \cdots \geq p^{(m)}$ be the upper processing time of jobs in $J\setminus\opt(I)$.
By \ref{online:worst-instance:pro:2} and \ref{online:worst-instance:pro:3} of \cref{lem:online:l1norm:worst-intance}, we know that \cref{alg:online:l1norm} has exactly $n-m$ untested jobs, i.e., $\abs{J\setminus\alg(I)}=n-m$.
Let $p_{A}^{(1)}\geq \cdots \geq p_{A}^{(n-m)}$ be the upper processing time of jobs in $J\setminus\alg(I)$.
Note that $(J\setminus\opt(I)) \cap (J\setminus\alg(I)) =\emptyset$ and $(J\setminus\opt(I)) \cup (J\setminus\alg(I)) =J$.
By \ref{online:worst-instance:pro:4} of \cref{lem:online:l1norm:worst-intance}, we know that only jobs in $J\setminus\opt(I)$ would contribute to the objective of the optimal solution. 
In contrast, all jobs in $J$ would contribute to the objective of the algorithm's solution.
This implies that 
\begin{equation}
    \opt = m \cdot p^{(m)} + \cdots + 1 \cdot p^{(1)}.
\label{equ:online:l1norm:opt}
\end{equation}
\cref{equ:online:l1norm:opt} provides the exact value of the optimal solution.
Now, we aim to seek an upper bound of $\alg$.
To do that, we reorder the jobs in $J$ so that jobs in $J\setminus\alg(I)$ and $J\setminus\opt(I)$ interleave.
If we reorder the jobs in the algorithm's solution, its total completion time can only become larger, providing an upper bound of $\alg$.
Formally, in the new order denoted by $\sigma$, jobs in $J\setminus\opt(I)$ will be put into odd positions ($1,3,\ldots,2m-1$) and jobs in $J\setminus\alg(I)$ will be put into even positions ($2,4,\ldots,2(n-m)$).
Note that there may exist some positions containing no jobs.
Let $f(\sigma)$ be the total completion time of job order $\sigma$.
Then, we have
\begin{align*}
    \alg\leq f(\sigma) &\leq 2 \cdot (n-m)\cdot p_{A}^{(n-m)} + \cdots + 2\cdot p_{A}^{(1)}+(2m-1)\cdot p^{(m)}+\cdots+1\cdot p^{(1)} \\
    &\leq 2 \cdot \left( (n-m)\cdot p_{A}^{(n-m)} + \cdots + 1 \cdot p_A^{(1)} \right) + 2 \cdot \left( m\cdot p^{(m)} +\cdots + 1\cdot p^{(1)} \right) \\
    &= 2 \cdot \left( (n-m)\cdot p_{A}^{(n-m)} + \cdots + 1 \cdot p_A^{(1)} \right) + 2 \cdot \opt . \tag*{[Due to \cref{equ:online:l1norm:opt}]}
\end{align*}

Now it remains to bound the first term of the above inequalities.
Let $\widetilde{I}$ be the offline instance constructed from $I$ by setting $\vpc=\mathbf{0}$.
Let $\cA(\widetilde{I})$ and $\opt(\widetilde{I})$ be the objective values of the algorithm's solution and optimal solution to instance $\widetilde{I}$, respectively.
By \cref{lem:online:l1norm:pc=0}, we have $\cA(\widetilde{I}) \leq (1+\epsilon) \opt(\widetilde{I})$ for any $\epsilon>0$.
Note that $\cA(\widetilde{I})=(n-m)\cdot p_{A}^{(n-m)} + \cdots + 1 \cdot p_A^{(1)}$ and $\opt(\widetilde{I})\leq\opt$.
Thus, we have 
\begin{align*}
    \alg&\leq 2 \cdot \left( (n-m)\cdot p_{A}^{(n-m)} + \cdots + 1 \cdot p_A^{(1)} \right) + 2 \cdot  \opt  \\
    &\leq 2\cdot (1+\epsilon) \cdot \opt + 2 \cdot \opt \\
    &= (4+2\epsilon) \cdot \opt.
\end{align*}
When $\vc=\mathbf{1}$, it is easy to verify that $\alg\leq 4\cdot \opt$ by By \cref{cor:online:l1norm:pc=0+c=1}.
\end{proofof}

\subsection{Algorithms for Auxiliary Instances}
\label{subsec:online:aux}

This subsection mainly shows two algorithms to solve the auxiliary instance $\wtI$.
Note that the auxiliary instance is a special case of the offline setting of our problem ($\vpc=\mathbf{0}$).
We first show a pseudo-polynomial algorithm that returns an optimal solution to $\wtI$ (See \cref{subsubsec:l1norm:online:pseudo}).
And then, we convert such an algorithm to obtain an FPTAS which proves \cref{lem:online:l1norm:pc=0} (See \cref{subsubsec:l1norm:online:FPTAS}).
Note that, by the hardness result stated in \cref{thm:online:l1norm:hardness}, FPTAS is the best possible algorithm for the auxiliary instance unless $\PP=\NP$.
We restate \cref{lem:online:l1norm:pc=0} for completeness.

\auxinstance*

Since $\pc_j=0$ for all $j\in J$ in the auxiliary instance, we use $p_j$ to denote $\ph_j$ for convenience.
Without loss of generality, we assume that $p_j\in\N_{\geq 0}$ for all $j\in J$.

\subsubsection{A Pseudo-polynomial Time Algorithm}
\label{subsubsec:l1norm:online:pseudo}

In this subsection, we mainly show that the auxiliary instance can be solved optimally by dynamic programming (DP) with a table of pseudo-polynomial size.
This proves the first part of \cref{lem:online:l1norm:pc=0}.
We first sort all jobs in non-increasing order by the value of $p_j$.
An exact dynamic programming algorithm is as follows.

\begin{definition}[DP Table]
The entry $D(C,j,k)$ stores the value of the minimum total budget used
among all solutions to a subinstance such that (\rom{1}) the subinstance contains the first $j$ jobs; (\rom{2}) the solution contains exactly $k$ tested jobs; (\rom{3}) the total completion time of the solution is at most $C$.
\end{definition}

Suppose that we know the value of $D(C,j,k)$ for all $C'\leq C$ and $k'\leq k$; we distinguish two cases to compute the value of $D(C,j+1,k)$.
If job $j+1$ is tested, then $j+1$ would not contribute to the objective; thus, $C$ remains the same.
Testing job $j+1$ would have a cost of $c_{j+1}+D(C,j,k-1)$.
If job $j+1$ is untested, then $j+1$ would contribute $p_{j+1}\cdot(j+1-k)$
to the total completion time.
Thus, the value of $D(C,j+1,k)$ is equal to $D(C-p_{j+1}\cdot(j+1-k),j,k)$ if $C-p_{j+1}\cdot(j+1-k)>0$.
Thus, the formula for the DP Table updating is as follows:

\begin{equation}
    D(C,j+1,k)= \min
    \begin{cases}
    c_{j+1}+D(C,j,k-1), &\text{if job $j+1$ is tested }\\
    D(C-p_{j+1}\cdot(j+1-k),j,k), &\text{if job $j+1$ is untested} 
    \end{cases}
\end{equation}
Note that $D(C,j,k)=\infty$ if $C<0$, i.e., the term $D(C-p_{j+1}\cdot(j+1-k),j,k)=\infty$ if $C<p_{j+1}\cdot(j+1-k)$.

\paragraph{Initial Cases} Note that $C\in\N_{\geq 0}$, $j,k\in [n]\cup\set{0}$.
Regarding the different faces of the DP table, we have the following three initial cases.
\begin{itemize}
    \item $D(C,j,0)=0$ for all $C\geq 1,j\in J$.
    \item $D(0,j,k)=\infty$ for all $j\in J, k\in[n]$.
    \item $D(C,1,k)$ has the following equality for all $C\geq 1$ and $k\in[n]$.
    \begin{equation*}
        D(C,1,k)=
        \begin{cases}
        0, &\text{if $p_1\leq C$} \\
        c_1, &\text{if $p_1 > C$}
        \end{cases}
    \end{equation*}
\end{itemize}

Let $P:=\max_{j\in J}\{p_j\}$ be the maximum processing time among all jobs.
Then, we have $C\leq n^2\cdot P$ where $n$ is the number of jobs.
Thus, the DP table has size $n^6\cdot P$, which implies that the running time of the DP above is pseudo-polynomial time.

Note that when $\vc=\mathbf{1}$, the optimal solution will would just simply test the first $B$ largest jobs.
This can be captured by the following observation (\cref{obs:l1norm:online:pc=0+c=1}).

\begin{observation}
Given an arbitrary instance $\cI:=(J,\vph,\vpc,\vc,B)$ with $\vpc=\mathbf{0}$ and $\vc=\mathbf{1}$, there exists an optimal solution such that it tests the last $B$ jobs after sorting jobs in non-decreasing order by the value of $p_j$.
\label{obs:l1norm:online:pc=0+c=1}
\end{observation}

\begin{proof}
Consider an arbitrary solution $S$, let $\sigma(S):=(q_1,\ldots,q_n)$ be the job order of the solution after sorting jobs in non-decreasing order by the processing time of jobs, where each entry $q_i$ is the processing time of $j_i$ in the solution.
Suppose that $S$ test $B'\leq B$ jobs.
Note that the first $B'$ of $\sigma(S)$ are $0$ and all the remaining entries are $p_j$, i.e., $q_i=0$ for all $i\in\set{1,\ldots,B'}$ and $q_i=p_i$ for all $i\in\set{B'+1,\ldots,n}$.
Let $O$ be the optimal solution that tests the last $B$ jobs after sorting jobs in non-decreasing order by the value of $p_j$.
It is easy to see that $\sigma(O) \preceq \sigma(S)$\footnote{Given two vectors $\vv=(v_1,\ldots,v_n),\vv'=(v_1',\ldots,v_n')$, $\vv \preceq \vv'$ means that $v_i\leq v_i'$ for all $i\in [n]$.} for any solution testing at most $B$ jobs.
This implies that $O$ always has an objective value no larger than any other feasible solution.
\end{proof}

\subsubsection{An FPTAS}
\label{subsubsec:l1norm:online:FPTAS}

To convert the above DP into an FPTAS, we need to eliminate the dependence on $P$ by reducing the number of distinct values in the $C$-columns of the DP table.


Before describing the rounding scheme below,
we need to know the optimal solution's maximum processing time (denoted by $P$).
We can assume that the algorithm knows the value of $P$.
This assumption can be removed by guessing the largest job in the optimal solution, which makes the algorithm loss an extra factor of $n$ on the running time.
The value of the largest job provides a lower bond of the optimal solution, i.e., $P\leq \opt$.

\paragraph{Rounding Scheme} Given an arbitrary auxiliary instance $\cI:=(J,(p_1,\ldots,p_n),\mathbf{0},\vc,B)$ and any $\epsilon>0$, let $\kappa=\frac{P\cdot\epsilon}{n^2}$ be the scaling parameter.
We construct a new instance $\cI':=(J',(p_1',\ldots,p_n'),\mathbf{0},\vc,B)$, where 
\begin{equation}
    p_j':=\left\lceil\frac{p_j}{\kappa}\right\rceil\cdot \kappa.
\label{equ:fptas:rounding}
\end{equation}
Note that the above inequality implies $p_j' \leq p_j + \kappa$.
After rounding the processing time of each job, job $i$ and $j$ would have the same processing time if $\frac{p_i}{\kappa}$ and $\frac{p_j}{\kappa}$ are between two identical adjacent integers.
Thus, the number of distinct processing times in $(p_1',\ldots,p_n')$ is at most $\frac{P}{\kappa}=\frac{n^2}{\epsilon}$.
Hence, the number of distinct values in the $C$-column of the DP table is at most $\frac{n^4}{\epsilon}$.
Therefore, when we apply the dynamic programming to the constructed instance, the DP table has a size of $\frac{n^8}{\epsilon}$.
This implies that the running time is $O(\text{poly}(n,\frac{1}{\epsilon}))$.

We conclude by proving the approximation ratio of this scheme.
Let $S\subseteq J$ be the set of remaining jobs produced by the dynamic programming, and $O\subseteq J$ be the optimal solution to instance $\cI$.
Let $S'\subseteq J'$ and $O'\subseteq J'$ be the corresponding jobs in $J'$ in the constructed instance.
We first sort all jobs in $S$ and $O$ by the value of processing time in non-decreasing order.
Note that the job order in $S$ (resp. $O$) and $S'$ (resp. $O'$) is identical according \cref{equ:fptas:rounding}.
Let $C_j$ be the completion time of job $j\in J$ and $C_j'$ be the completion of job $j'\in J'$.
Since we round up the processing time of each job, we have $\alg=\sum_{j\in S}C_j \leq \sum_{j\in S'}C_j'$.
Since $S$ is an optimal solution to instance $\cI'$, we have $\sum_{j\in S'}C_j' \leq \sum_{j\in O'}C_j'$.
Then, we have the following inequalities:

\begin{align*}
\alg 
&=\sum_{j\in S}C_j \\
&\leq \sum_{j\in S'}C_j' \tag*{[Due to \cref{equ:fptas:rounding}]}\\
&\leq \sum_{j\in O'}C_j' \tag*{[Due to $S'$ is optimal to $\cI'$]} \\
&= \abs{O}\cdot p_1' + (\abs{O}-1)\cdot p_2' + \cdots \\
&\leq \abs{O}\cdot (p_1+\kappa) + (\abs{O}-1)\cdot (p_2+\kappa) + \cdots \tag*{[Due to $p_j' \leq p_j + \kappa$]}\\
&\leq \opt + n^2 \cdot \kappa \tag*{[Due to $\abs{O} \leq n$]} \\
&=\opt +\epsilon \cdot P \tag*{[Due to $\kappa=\frac{P\cdot \epsilon}{n^2}$]} \\
&\leq (1+\epsilon)\cdot \opt \tag*{[Due to $P\leq \opt$]}
\end{align*}

According to \cref{lem:online:l1norm:pc=0}, we have the following corollary easily.
\begin{corollary}\label{cor:online:l1norm:pc}
    Consider an arbitrary instance $\cI:=(J,\vph,\vpc,\vc,B)$ with all the $\pc_j$'s are the same. By reducing it to the instance $\wtI:=(J,\widetilde{\vph},\0,\vc,B)$ where $\widetilde{\ph_j}=\ph_j-\pc_j$ for each $j\in J$, an FPTAS can be obtained.
\end{corollary}

\subsection{Properties of Worst-case Instance}
\label{subsec:online:worst-instance}

In this subsection, we mainly show some properties of the worst-case instance of \cref{alg:online:l1norm}.
For completeness, we restate \cref{lem:online:l1norm:worst-intance}.

\worstinstance*

\begin{proof}
Given an arbitrary instance $\cI$ with job set $J$ and budget $B$, we show how to modify $\cI$ such that $I$ satisfies \ref{online:worst-instance:pro:2}, \ref{online:worst-instance:pro:3}, \ref{online:worst-instance:pro:4}, and keep the competitive ratio of the constructed instance non-decreasing simultaneously (satisfying \ref{online:worst-instance:pro:1}).

\paragraph{Satisfying \ref{online:worst-instance:pro:2}} We first show that the competitive ratio of the instance is non-decreasing after making $\cI$ satisfy \ref{online:worst-instance:pro:2}.
In the case where $\alg(\cI) \cap \opt(\cI)=\emptyset$, the instance naturally satisfies \ref{online:worst-instance:pro:1} and \ref{online:worst-instance:pro:2}. 
If it is not in this case, we define $S:=\alg(\cI) \cap \opt(\cI)\ne\emptyset$.
Now, we construct an instance $\cI^1$ with job set $J^1$ and budget $B^1$ where $B^1:=B-\sum_{j\in S}c_j$.
For each job $j$ in $J$, we have one job $j^1$ in $J^1$.
Let $S^1 \subseteq J^1$ be the corresponding job set $S\subseteq J$.
For each job $j\in J$, the corresponding job $j^1$ has the same lower processing time and upper limit, i.e., $\pc_{j^1}:=\pc_j$ and $\ph_{j^1}:=\ph_j$.
For each job $j\in J\setminus S$, the corresponding job $j^1$ has the same testing cost, i.e., $c_{j^1}:=c_{j}$.
For each job $j\in S$, the corresponding job $j^1$ has no testing cost, i.e., $c_{j^1}:=0$.
To prove the instance $\cI^1$ maintains \ref{online:worst-instance:pro:1}, we show that $\com(\opt(\cI))=\com(\opt(\cI^1))$ and $\com(\cI,\alg(\cI))=\com(\cI^1,\alg(\cI^1))$.
Since each job $j$ and its corresponding job $j^1$ has the same lower processing time and upper limit, we only need to prove that $\alg(\cI^1)=\alg(\cI)$ and $\opt(\cI^1)=\opt(\cI)$, i.e., both the algorithm's and optimal solution remain the same.
We prove these two equations in \cref{clm:online:worst-instance:I-1-alg} and \cref{clm:online:worst-instance:I-1-opt} separately.

\begin{claim}
Let $\alg(\cI)$ and $\alg(\cI^1)$ be the solution returned by \cref{alg:online:l1norm} when the input is $\cI$ and $\cI^1$, respectively.
Then, we have $\alg(\cI)=\alg(\cI^1)$.
\label{clm:online:worst-instance:I-1-alg}
\end{claim}

\begin{proof}[Proof of \cref{clm:online:worst-instance:I-1-alg}]
Note that $\alg(\cI)$ and $\alg(\cI^1)$ is an optimal solution to instance $(J,\vph,\0,\vc,B)$ and $(J^1,\vph,\0,\vc^1,B^1)$, respectively.
Let $Z\subseteq J^1$ be the corresponding job set to $\alg(\cI)$. 
To prove \cref{clm:online:worst-instance:I-1-alg}, we only need to show that $Z$ is an optimal solution to instance $(J^1,\vph,\0,\vc^1,B^1)$.
Firstly, it is easy to verify that $Z$ is a feasible solution to $\cI^1$ since $\sum_{j^1\in Z}c_{j^1}=B^1$.
Assume the contrary that $Z$ is not an optimal solution to $\cI^1$, then there must exist another solution $Z'$ such that $\com(\cI^1,Z')<\com(\cI^1,Z)$.
Note that $\com(\cI,\alg(I))=\com(\cI^1,Z)$.
Without loss of generality, we assume that $S^1 \subseteq Z'$ since each job in $S^1$ has no testing cost.
Let $Y\subseteq J$ be the corresponding job set to $Z'$.
Note that $\com(\cI,Y)=\com(\cI^1,Z')$.
Since $Z'$ is a feasible solution to $\cI^1$, we have $\sum_{j^1\in Z'}c_{j^1}\leq B^1$.
Thus, we have $\sum_{j\in Y}c_j \leq B$ which implies that $Y$ is a feasible solution to $\cI$.
Therefore, we have $\com(\cI,Y)<\com(\cI,\alg(\cI))$ which contradicts the optimality of our algorithm.
\end{proof}

\begin{claim}
There exists an optimal solution $\opt(\cI^1)$ to instance $\cI^1$ such that $\opt(\cI^1)=\opt(\cI)$.
\label{clm:online:worst-instance:I-1-opt}
\end{claim}

\begin{proof}[Proof of \cref{clm:online:worst-instance:I-1-opt}]
Let $O\subseteq J^1$ be the corresponding job set to $\opt(\cI)$.
To prove \cref{clm:online:worst-instance:I-1-opt}, we only need to show that $O$ is an optimal solution to instance $\cI^1$.
Firstly, it is easy to verify that $O$ is a feasible solution $\cI^1$ since $\sum_{j\in O}c^1_j = B^1$.
Assume the contrary that $O$ is not an optimal solution to $\cI^1$, then there must exist another solution $O'$ such that $\com(\cI^1,O')<\com(\cI^1,O)$.
Note that $\com(\cI,\opt(\cI))=\com(\cI^1,O)$.
Without loss of generality, we assume that $S^1\subseteq O'$ since each job in $S^1$ has no testing cost.
Let $Q\subseteq J$ be the corresponding job set to $O'$.
Note that $\com(\cI,Q)=\com(\cI^1,O')$.
It is easy to verify that $Q$ is a feasible solution $\cI$.
Thus, we have $\com(\cI,Q)<\com(\cI,\opt(\cI))$ which contradicts the optimality of $\opt(\cI)$.
\end{proof}

\paragraph{Satisfying \ref{online:worst-instance:pro:3}}
We now convert instance $\cI^1$ to $\cI^2$ such that $\cI^2$ satisfies \ref{online:worst-instance:pro:2}, \ref{online:worst-instance:pro:3} and keep the competitive ratio of $\cI^2$ non-decreasing simultaneously.
From the above analysis, we know that $\cI^1$ satisfies \ref{online:worst-instance:pro:2}.
In the case where $\alg(\cI^1)\cup\opt(\cI^1)=J^1$, the instance naturally satisfies \ref{online:worst-instance:pro:1}, \ref{online:worst-instance:pro:2} and \ref{online:worst-instance:pro:3}.
If it is not in this case, we define $W^1:=J^1\setminus \left( \alg(\cI^1) \cup \opt(\cI^1) \right)\ne\emptyset$.
Now, we construct an instance with job set $J^2$ and budget $B^2$ where $B^2:=B^1$.
For each job $j^1\in J^1$, we have one job $j^2$ in $J^2$.
Let $W^2 \subseteq J^2$ be the corresponding job set $W^1 \subseteq J^1$.
For each job $j^1\notin W^1$, the corresponding job $j^2$ has the same lower processing time, upper limit and the testing cost, i.e., $\pc_{j^2}:=\pc_{j^1}$, $\ph_{j^2}:=\ph_{j^1}$ and $c_{j^2}:=c_{j^1}$. 
For each job $j^1\in W^1$, we set the parameters of the corresponding job $j^2$ as follows: (\rom{1}) $\pc_{j^2}=\ph_{j^2}:=\ph_{j^1}$; (\rom{2}) $c_{j^2}:=0$.
To prove the instance $\cI^2$ maintains \ref{online:worst-instance:pro:3}, we show that (\rom{1}) the corresponding job set of $\alg(\cI^1)$ is a job set that returned by \cref{alg:online:l1norm} when the input is $\cI^2$;
(\rom{2}) there exists an optimal solution $\opt(\cI^2)$ to $\cI^2$ such that $\opt(\cI^2)=W^2\cup L$, where $L$ represents the corresponding job set of $\opt(\cI^1)$ in $J^2$.
To prove the instance $\cI^2$ maintains \ref{online:worst-instance:pro:1}, we show that (\rom{3}) $\com(\opt(\cI^1))=\com(\opt(\cI^2))$; (\rom{4}) $\com(\alg(\cI^1))=\com(\alg(\cI^2))$.
Since \cref{alg:online:l1norm} only accesses the upper processing time and testing cost of each job, the proof of \cref{clm:online:worst-instance:I-1-alg} still works for the current case.
Thus, we have (\rom{1}) and (\rom{3}).
In the following, we prove $\opt(\cI^2)=W^2\cup L$ in
\cref{clm:online:worst-instance:I-2-opt}.
Note that, for each job $j^2\in W^2$, $\pc_{j^2}=\ph_{j^2}:=\ph_{j^1}$ and $c_{j^2}=0$.
Thus, $\com(\opt(\cI^1))=\com(\opt(\cI^2))$.
Now, we focus on \cref{clm:online:worst-instance:I-2-opt}.

\begin{claim}
There exists an optimal solution to instance $\cI^2$, denoted by $\opt(\cI^2)$, such that $L\cup W^2=\opt(\cI^2)$, where $L$ represents the corresponding job set of $\opt(\cI^1)$ in $J^2$.
\label{clm:online:worst-instance:I-2-opt}
\end{claim}

\begin{proof}[Proof of \cref{clm:online:worst-instance:I-2-opt}]
To prove \cref{clm:online:worst-instance:I-2-opt}, we only need to show that $L\cup W^2$ is an optimal solution to instance $\cI^2$.
Firstly, it is easy to verify that $L\cup W^2$ is a feasible solution to $\cI^2$.
Assume the contrary that $L\cup W^2$ is not an optimal solution to $\cI^2$, then there must exist another solution $U\subseteq J^2$ such that $\com(\cI^2,U)<\com(\cI^2,L\cup W^2)$.
Note that $\com(\cI^2,L\cup W^2)=\com(\cI^1,\opt(\cI^1))$.
Without loss of generality, we assume that $W^2 \cap U =\emptyset$ since each job in $W^2$ has no testing cost and $\pc_{j^2}=\ph_{j^2}$ for all $j^2\in W^2$.
Let $U^1\subseteq J^1$ be the corresponding job set to $U$.
It is not hard to see that $U^1$ is a feasible solution to $\cI^1$ and $\com(\cI^1,U^1)=\com(\cI^2,U)$.
Thus, we have $\com(\cI^1,U^1)<\com(\cI^1,\opt(\cI^1))$ which contradicts the optimality of $\opt(\cI^1)$.
\end{proof}

\paragraph{Satisfying \ref{online:worst-instance:pro:4}}

We now covert instance $\cI^2$ to $\cI^3$ such that $\cI^3$ satisfies \ref{online:worst-instance:pro:2}, \ref{online:worst-instance:pro:3}, \ref{online:worst-instance:pro:4} and keep the competitive ratio of $\cI^3$ non-decreasing simultaneously.
From the above analysis, we know that $\cI^2$ satisfies \ref{online:worst-instance:pro:2} and \ref{online:worst-instance:pro:3}.
In the case where $\pc_{j^2}=\ph_{j^2}$ for all $j^2\in\alg(\cI^2)$ and $\pc_{j^2}=0$ for all $j^2\in\opt(\cI^2)$, the instance naturally satisfies \ref{online:worst-instance:pro:2}, \ref{online:worst-instance:pro:3} and \ref{online:worst-instance:pro:4}.
If it is not in this case, we construct an instance $\cI^3$ with job set $J^3$ and budget $B^3$ where $B^3:=B^2$.
For each job $j^2$ in $J^2$, we have one job $j^2$ in $J^2$.
For each job $j^2\in \opt(\cI^2)$, the corresponding job $j^3$ has zero lower processing, i.e., $\pc_{j^3}:=0$, $\ph_{j^3}:=\ph_{j^2}$ and $c_{j^3}:=c_{j^2}$.
For each job $j^2\in \alg(\cI^2)$, we set the parameters of the corresponding job $j^3$ as follows: $\pc_{j^3}=\ph_{j^3}:=\ph_{j^2}$ and $c_{j^3}:=c_{j^2}$.
To prove the instance $\cI^3$ maintains \ref{online:worst-instance:pro:4}, we show that $\com(\opt(\cI^3))\leq \com(\opt(\cI^2))$ and $\com(\alg(\cI^3)) = \com(\alg(\cI^2))$.
Since \cref{alg:online:l1norm} only accesses the upper processing time and testing cost of each job, the proof of \cref{clm:online:worst-instance:I-1-alg} still works for the current case.
Thus, we have $\com(\alg(\cI^3)) = \com(\alg(\cI^2))$.
Since $\vpc_{3} \preceq \vpc_{2}$ and $\vph_{3} \preceq \vph_{2}$, we have $\com(\opt(\cI^3))\leq \com(\opt(\cI^2))$.
Thus, the constructed instance $\cI^3$ satisfies \ref{online:worst-instance:pro:1}, \ref{online:worst-instance:pro:2}, \ref{online:worst-instance:pro:3} and \ref{online:worst-instance:pro:4}.
\end{proof}

\section{\SLPB/ under Makespan Minimization}
\label{sec:makespan}

\subsection{Offline Setting}

Getting the optimal offline solution is straightforward when the objective is makespan minimization.
The problem is equivalent to the classical knapsack problem.
The following theorem can capture this.

\begin{theorem}
The problem \SLPBm/ is equivalent to the classical knapsack problem.
\label{thm:offline:makespan:equal}
\end{theorem}

\begin{proof}
To prove the theorem, we show the following two reduction directions: (\rom{1}) there is a polynomial time reduction from \SLPBm/ to knapsack; (\rom{2}) there is a polynomial time reduction from knapsack to \SLPBm/.
These two reductions are similar and from the classical knapsack problem.
We first define the knapsack problem in the following.
A knapsack instance consists of an item set $N:=\set{1,\ldots,n}$ and the capacity of the knapsack $C$.
Each item is associated with a value $v_i$ and a weight $w_i$.
The goal is to select an item set $N'\subseteq N$ such that $\sum_{i\in N'}w_i\leq C$ and $\sum_{i\in N'}v_i$ is maximized.

\paragraph{\SLPBm/ $\leqp$ Knapsack}
In this direction, we show that \SLPBm/ is a special case of the knapsack problem.
Given an arbitrary instance $\cI=(J,\vph,\vpc,\vc,R)$ of \SLPBm/, we construct a knapsack instance $\cK=(N,\vv,\vw,C)$ as follows.
For each job $j\in J$, we have one item $i$ in $N$.
The value of item $i$ is defined as $v_i:=\pc_j-\ph_j$, and the weight is defined as $w_i:=c_j$.
The capacity of the knapsack is $C:=R$.
Now, we show that if a polynomial time algorithm $\alg$ solves the knapsack problem, then $\alg$ solves \SLPBm/.
To this end, we only need to prove that the optimal solution to the constructed knapsack instance is also optimal for the original \SLPBm/ instance.

Given a job set $J'\subseteq J$, let $\pc(J'):=\sum_{j\in J'}\pc_j$ and $\ph(J'):=\sum_{j\in J'}\ph_j$.
Given an arbitrary instance $\cI$ of \SLPBm/, let $S$ be an arbitrary feasible solution and $F(S,\cI)$ be the objective value of the solution $S$, i.e., $\sum_{j\in S}c_j\leq R$ and $F(S,\cI)=\pc(S)+\ph(J\setminus S)$.
Thus, we have $F(S,\cI)=\ph(j)-(\pc(S)-\ph(S))$.
It is not hard to see that the optimal solution to the knapsack instance maximizes the value of $\pc(S)-\ph(S)$.
Therefore, it is also an optimal solution to the original \SLPBm/ instance.

\paragraph{Knapsack $\leqp$ \SLPBm/}
In this direction, we show that the knapsack problem is a special case of \SLPBm/.
Given an arbitrary instance of the knapsack problem $\cK=(N,\vv,\vw,C)$, we construct an instance of \SLPBm/ $\cI=(J,\pc,\ph,\vc,R)$ as follows.
For each item $i\in N$, we have one job in $J$.
The upper limit of job $j$ is defined as $\ph_j:=v_i$.
And all  processing times are equal to $0$, i.e., $\pc_j:=0$ for all $j\in J$.
The test cost of job $j$ is defined as $c_j:=w_i$, and the total testing budget is defined as the capacity of the knapsack $R:=C$.
Then, by the same argument used in the previous reduction, we know that the optimal solution to the constructed \SLPBm/ instance is the same as the optimal solution to the original knapsack instance.

\end{proof}

By \cref{thm:offline:makespan:equal}, we have the following two corollaries from the results of the classical knapsack problem.

\begin{corollary}
The problem \SLPBm/ with makespan minimization objective is $\NP$-hard, even for instances $\cI=(J,\ph,\0,\vc,R)$.
\label{cor:makespan:offline:hardness}
\end{corollary}

\begin{corollary}
The problem \SLPBm/ with makespan minimization objective admits a pseudo-polynomial time algorithm.
\label{cor:makespan:offline:pseudo}
\end{corollary}

One can easily transfer the above pseudo-polynomial time algorithm to an FPTAS using the same argument of the knapsack's algorithm.
Thus, we have the following corollary.

\begin{corollary}
The problem \SLPBm/ with makespan minimization objective admits an FPTAS.
\label{cor:makespan:offline:FPTAS}
\end{corollary}

When the testing cost of each job is unit, the optimal solution will test the first $K$ jobs with the largest $\ph-\pc$ (See \cref{alg:offline:makespan}).
Thus, we have the following simple observation.

\begin{algorithm}[htb]
\caption{\textsc{Offline $\ell_{\infty}$-norm Minimization}}
\label{alg:offline:makespan}
\begin{algorithmic}[1]
\REQUIRE  processing time vector $\vpc\in\N_{\geq 0}^n$; Upper limit processing time vector $\vph\in\N_{\geq 0}^n$; Testing budget $K\leq n$.
\ENSURE A testing job set $S$.
\FOR{every job $j\in J$}
\STATE $d_j\leftarrow \ph_j-\pc_j$. 
\ENDFOR
\STATE Sort all jobs in non-increasing order by the value of $d_j$.
\STATE Let $S$ be the first $K$ jobs.
\RETURN The job set $S$.
\end{algorithmic}
\end{algorithm}

\begin{observation}
\cref{alg:offline:makespan} finds the optimal offline solution when the goal is to minimize the makespan.
\label{thm:offline:makespan:ratio}
\end{observation}

\begin{proof}
Given a job set $J'$, let $\ph(J')=\sum_{j\in J'}\ph_j$ and $\pc(J')=\sum_{j\in J'}\pc_j$.
Given an arbitrary instance $\cI$, let $S$ be an arbitrary feasible solution and $F(S,\cI)$ be the objective value of the solution $S$, i.e., $\abs{S}\leq K$ and $F(S,\cI)=\pc(S)+\ph(J\setminus S)$.
Thus, we have $F(S,I)=\ph(J)-(\ph(S)-\pc(S))$.
Clearly, the optimal solution will be a job set $S$ such that $\ph(S)-\pc(S)$ is maximized since $\ph(J)$ is a fixed number.
Thus, \cref{alg:offline:makespan} returns an optimal solution.
\end{proof}

\subsection{Oblivious Setting}

In this section, we consider the oblivious version of the problem.
Recall that the  processing time vector $\vpc$ is the only hidden information for the algorithm.
We first give the upper and lower bound for the uniform testing cost case where each job has the same testing cost, and then present the results for the general case in the next subsection.

\subsubsection{Uniform Testing Cost Variant}

We first give a lower bound of the oblivious setting and then show that a simple greedy algorithm is essentially optimal.

\begin{theorem}
There is no deterministic algorithm whose competitive ratio is better than $2$, even for the setting of uniform testing cost.
\label{thm:online:makespan:hardness}
\end{theorem}

\begin{proof}
We prove the theorem by giving the following hard instance $\cI$.
The instance consists of $n$ jobs and $K=\frac{n}{2}$.
All jobs have the same upper limits on the processing time $1$, i.e., $\vph=\mathbf{1}$.
Let $\alg$ be an arbitrary algorithm, and $S$ be the solution returned by $\alg$.
The adversary can always make the testing operation of $\alg$ useless.
Namely, the adversary can set up an instance such that $\pc_j=0$ for all $j\in J\setminus S$ and $\pc_j=1$ for all $j\in S$.
Since $K=\frac{n}{2}$, there always exists an instance $\cI$ such that $F(S,\cI)=n$ for any algorithms, while the optimal solution has value $\opt(\cI)=\frac{n}{2}$.
Thus, any algorithms have a competitive ratio of at least $2$.
\end{proof}

Now, we give a simple greedy algorithm that is essentially optimal.
The greedy algorithm (\cref{alg:online:makespan}) first sorts all jobs in non-increasing order by their $\ph_j$, then tests the first $K$ jobs.

\begin{algorithm}[htb]
\caption{\textsc{Oblivious Makespan Minimization}}
\label{alg:online:makespan}
\begin{algorithmic}[1]
\REQUIRE Upper limit processing time vector $\vph\in\N_{\geq 0}^n$; Testing budget $K\leq n$.
\ENSURE A testing job set $S$.
\STATE Sort all jobs in non-increasing order by the value of $\ph_j$.
\STATE Let $S$ be the first $K$ jobs.
\RETURN The job set $S$.
\end{algorithmic}
\end{algorithm}

\begin{theorem}
\cref{alg:online:makespan} is a $2$-competitive algorithm.
\label{thm:online:makespan:uniform:ratio}
\end{theorem}

\begin{proof}
Let $S$ be the job set returned by \cref{alg:online:makespan} and $S^*$ be the jobs that are tested by the optimal solution.
We define $\bar{S}:=S\setminus S^*$ for notation convenience.
Given a job set $J'$, let $\alg(J')$ and $\opt(J')$ be the total processing time of jobs in $J'$ in the solution returned by \cref{alg:online:makespan} and the optimal solution, respectively.
Then, we have $\alg=\alg(\bar{S})+\alg(J\setminus\bar{S})$.
In the following, we bound $\alg(\bar{S})$ and $\alg(J\setminus\bar{S})$ by $\opt$, respectively.

\paragraph{Jobs in $\bar{S}$} 
Note that $\bar{S}$ is a set of jobs that \cref{alg:online:makespan} tests, but the optimal solution does not.
Thus, for each job in $\bar{S}$, the processing time in $\alg$'s schedule is no larger than its processing time in $\opt$'s schedule. 
Thus, we have $\alg(\bar{S})\leq\opt(\bar{S})\leq\opt$.

\paragraph{Jobs in $J\setminus\bar{S}$} 
Let $\widetilde{S}:=S\cap S^*$. 
Then, we have $\alg(J\setminus\bar{S})=\alg(\widetilde{S})+\alg(J\setminus S)$.
Note that $\widetilde{S}$ is a set of jobs that are tested in both $\alg$'s schedule and $\opt$'s schedule.
Thus, we have $\alg(\widetilde{S})=\opt(\widetilde{S})$.
Given a job set $J'$, let $\ph(J')=\sum_{j\in J'}\ph_j$.
Then, we have $\alg(J\setminus S)=\ph(J\setminus S)$ and $\opt(J\setminus S^*)=\ph(J\setminus S^*)$.
Since \cref{alg:online:makespan} greedily picks the first $K$ jobs with largest $\ph$, we have $\ph(S)\geq\ph(S^*)$.
This implies $\ph(J\setminus S)\leq \ph(J\setminus S^*)$.
Hence, we have $\alg(J\setminus S)\leq \opt(J\setminus S^*)$.
Therefore, we have:
$$
\alg(J\setminus\bar{S})=\alg(\widetilde{S})+\alg(J\setminus S)\leq \opt(\widetilde{S})+\opt(J\setminus S^*) \leq \opt.
$$

Thus, combining the above cases, we have $\alg=\alg(\bar{S})+\alg(J\setminus\bar{S})\leq 2\cdot\opt$.
\end{proof}

\subsubsection{$(2+\epsilon)$-competitive Algorithm}

Now we consider the general case. 
The basic idea of the algorithm is the same as \cref{alg:online:l1norm}.
Define an {\em auxiliary instance} which is the same as \cref{def:aux-instance}.
By \cref{cor:makespan:offline:FPTAS}, we know that there is an FPTAS for the auxiliary instance.
Thus, we get a $(1+\epsilon)$-approximation solution for the auxiliary instance.
In the following, we show that \cref{alg:online:l1norm} computes a $(2+\epsilon)$-approximation solution to \SLPBm/ with makespan minimization objective.
The analysis framework is based on the property of the worst-case instance stated in \cref{lem:online:l1norm:worst-intance}, which is similar to the proof of \cref{thm:online:l1norm:ratio}.
We restate \cref{lem:online:l1norm:worst-intance} for completeness.
It is not hard to see that the proof of \cref{lem:online:l1norm:worst-intance} still works even if we change the objective from the total completion time to the makespan.

\worstinstance*

Now, we are ready to prove the following theorem.

\begin{theorem}
Given an arbitrary instance $\cI:=(J,\vph,\vph,\vc,R)$ with $\vpc$ is hidden for the algorithm, \cref{alg:online:l1norm} returns a solution that is $(2+\epsilon)$-approximated when the objective is makespan minimization.
\label{thm:online:makespan:non-uniform:ratio}
\end{theorem}

\begin{proof}
Note that \cref{lem:online:l1norm:worst-intance} proved that the worst-case instance of \cref{alg:online:l1norm} must satisfy the following two properties: (\rom{1}) the optimal solution tests exactly the jobs that \cref{alg:online:l1norm} does not test;
(\rom{2}) all jobs $j$ that are tested by the optimal solution have $\pc_j=0$ and all jobs $j$ that are tested by \cref{alg:online:l1norm} have $\pc_j=\ph_j$.
Let $S$ and $S^*$ be the job set tested by \cref{alg:online:l1norm} and the optimal solution, respectively.
Note that $S\cup S^*=J$ and $S\cap S^*=\emptyset$ by \cref{lem:online:l1norm:worst-intance}.
Let $\alg$ and $\opt$ be the objective value of the algorithmic and optimal solution.
By \cref{lem:online:l1norm:worst-intance}, we have $\alg=\sum_{j\in S}\pc_j+\sum_{j\in S^*}\ph_j$ and $\opt=\sum_{j\in S}\pc_j$.
Observe that if we have the following claim, then \cref{alg:online:l1norm} is a $(2+\epsilon)$-approximation algorithm.

\begin{claim}
$\sum_{j\in S^*}\ph_j \leq (1+\epsilon)\cdot \opt$.
\end{claim}

The above claim is true since \cref{alg:online:l1norm} returns a $(1+\epsilon)$-approximation solution for the auxiliary instance.
\end{proof}

\end{document}